\documentclass[12pt]{iopart}
\usepackage{gensymb}
\usepackage{graphicx}

\usepackage{subcaption}
\usepackage{amsbsy}
\usepackage{color}
\usepackage{textcomp}
\usepackage{float}
\usepackage{sidecap}
\usepackage{siunitx}
\usepackage{amssymb}
\usepackage{wrapfig}
\expandafter\let\csname equation*\endcsname\relax

\expandafter\let\csname endequation*\endcsname\relax

\usepackage{lineno}
\usepackage{amsmath}
\usepackage{makecell}
\usepackage{upgreek}

\usepackage{etoolbox} %% <- for \cspreto, \csappto

\newcommand*\linenomathpatch[1]{%
  \cspreto{#1}{\linenomath}%
  \cspreto{#1*}{\linenomath}%
  \csappto{end#1}{\endlinenomath}%
  \csappto{end#1*}{\endlinenomath}%
}

\linenomathpatch{equation}
\linenomathpatch{gather}
\linenomathpatch{multline}
\linenomathpatch{align}
\linenomathpatch{alignat}
\linenomathpatch{flalign}

\protected\def\PRE{\ifmmode \mathrm{PRE} \else PRE\fi}

\usepackage{ulem}
\newcommand{\stkout}[1]{\ifmmode\text{\sout{\ensuremath{#1}}}\else\sout{#1}\fi}
\newcommand{\Edit}[1]{{\color{black}#1}}
\newcommand{\DEdit}[1]{\iffalse{\color{red}\stkout{#1}}\fi}

\begin{document}
%\linenumbers

\title{A predictive formula for the H-mode electron separatrix density: Bridging regression and physics-based models across C-Mod, AUG and JET tokamaks}

\author{D. Silvagni${^{1}}$, O. Grover${^{1}}$, A. Stagni${^2}$, J. W. Hughes${^3}$, M. A. Miller${^1}$, B. Lomanowski${^{4}}$, L. Balbinot${^{5}}$, G. Ciraolo${^6}$, W. Dekeyser${^7}$, M. Dunne${^{1}}$, L. Frassinetti${^8}$, C. Giroud${^{9}}$, T. Happel${^{1}}$, I. Jepu${^{9}}$, A. Kallenbach${^{1}}$, A. Kirjasuo${^{10}}$, A. Kuang$^{11}$, T. Luda${^{1}}$, D. Moulton${^{9}}$, O. Pan${^{1}}$, C. Perez von Thun${^{12}}$, T. P{\"u}tterich${^{1}}$, G. Rubino${^{13}}$, S. A. Silburn${^{9}}$, H. J. Sun${^{9}}$, D. Umezaki${^{14}}$, H. Zohm${^{1}}$, the ASDEX Upgrade team${^{a}}$, JET contributors${^{b}}$ and the EUROfusion tokamak exploitation team${^{c}}$}

\address{
$^1$Max-Planck-Institut f{\"u}r Plasmaphysik, 85748 Garching, Germany \\
$^2$Consorzio RFX (CNR, ENEA, INFN, Universit\`a di Padova, Acciaierie Venete SpA), Corso Stati Uniti 4, 35127 Padova, Italy \\ 
$^3$MIT Plasma Science and Fusion Center, Cambridge, MA, USA \\ 
$^4$Oak Ridge National Laboratory, Oak Ridge, TN 37831-6169, USA \\ 
$^5$DTT S.c. a r.l., 00044 Frascati, Roma, Italy \\
$^6$IRFM, CEA, St. Paul-Lez-Durance, France \\ 
$^7$KU Leuven, Department of Mechanical Engineering, 3001 Leuven, Belgium \\ 
$^8$KTH Royal Institute of Technology, Division of Electromagnetic Engineering and Fusion Science, Stockholm, Sweden \\ 
$^{9}$UKAEA, Culham Science Centre, Abingdon, OX14 3DB, UK \\ 
$^{10}$VTT, PO Box 1000, FI-02044 VTT Espoo, Finland \\  
$^{11}$Commonwealth Fusion Systems, Devens, Massachusetts 01434, USA \\ 
$^{12}$Institute of Plasma Physics and Laser Microfusion, Hery 23, 01-497 Warsaw, Poland \\ 
$^{13}$Istituto per la Scienza e la Tecnologia dei Plasmi, CNR, Bari, Italy \\
$^{14}$National Institutes for Quantum Science and Technology, Naka, Ibaraki 311-0193, Japan \\
$^{\mathrm{a}}$See author list of H. Zohm, 2024 Nucl. Fusion 64 112001 \\ 
$^{\mathrm{b}}$See author list of C. F. Maggi, 2024 Nucl. Fusion 64 112012 \\ 
$^{\mathrm{c}}$See author list of E. Joffrin, 2024 Nucl. Fusion 64 112019
}
 
\ead{davide.silvagni@ipp.mpg.de}
\vspace{10pt}

%\begin{indented}

%\item[]June 2021

%\end{indented}

\begin{abstract}
The electron density at the separatrix ($n_{e,\mathrm{sep}}$) plays a central role in balancing energy confinement, detachment achievement, and ELM suppression in tokamaks, thereby influencing core-edge integration.
To study what determines this key parameter, a database of H-mode separatrix density measurements from Alcator C-Mod, ASDEX Upgrade, and JET tokamaks has been assembled using a consistent analysis method across all devices. This dataset is used to derive a regression scaling expression based solely on engineering parameters, and the results are compared to predictions from the two-point model. The agreement found is remarkable: both the regression and model provide similar parameter dependencies and tokamak-specific multiplicative constants. 
Building on this agreement, a fully predictive formula that combines the regression dependencies and the two-point model multiplicative constant is proposed. This formula is able to estimate $n_{e,\mathrm{sep}}$ across the three machines within a factor of 1.5, and provides projections to next-step devices (ITER, SPARC, DTT, JT-60SA and COMPASS-U) that are in agreement with available SOLPS simulations.

\end{abstract}

%\submitto{\NF}
%\maketitle

\section{Introduction}

One of the key open challenges on the pathway to building a fusion power plant based on the tokamak concept is the determination of a so-called core-edge integrated plasma scenario. In such a scenario, the plasma needs to combine high fusion energy production with acceptable heat and particle loads on first-wall materials, the latter usually achieved via pronounced detachment~\cite{kallenbach_2015_nuclfusion}. Moreover, transient heat loads induced by edge-localized modes (ELMs) should be avoided~\cite{Eich_2017}, along with plasma disruptions. 
The electron density at the separatrix $n_{e,\mathrm{sep}}$ is a key parameter that influences the above-mentioned processes: It affects H-mode confinement by altering the pedestal stability~\cite{Dunne_2017_hfshd, Frassinetti_2023, Hughes_2011, Osborne_2015, Silvagni_2024}, detachment achievement by influencing the required impurity concentration needed to detach~\cite{Henderson_2021, Kallenbach_2016, Goldston_2017}, access to no-ELM regimes~\cite{Harrer_2018, Miller_2025, Suttrop_2018, Silvagni_2023}, and it poses an upper limit to H-mode operation (the so-called density limit)~\cite{Lipschultz_1984, Maingi_2005, Eich_2021}. Therefore, reliable, quantitative predictions of $n_{e,\mathrm{sep}}$ are essential for evaluating and
designing next-step fusion experiments and reactors. 

Several previous studies highlighted the connection between the (upstream) separatrix density and the (downstream) divertor target conditions, via the so-called two-point model~\cite{Stangeby_2018, Krasheninnikov_2016}. At DIII-D~\cite{Leonard_2017}, JET~\cite{Lomanowski_2023} and WEST~\cite{Bourdelle_2023}, $n_{e,\mathrm{sep}}$ has been shown to be clearly linked to the measured electron temperature in the divertor region, as expected from the two-point model equations. In ASDEX Upgrade (AUG), a strong connection between the electron separatrix density and the sub-divertor neutral pressure $p_{0,\mathrm{div}}$ has been observed both experimentally~\cite{Kallenbach_2018} and in simulations~\cite{Kallenbach_2019}. This finding has been reconciled with the two-point model by relating the target ion flux to the divertor neutral pressure. More recently, similar relationships connecting $n_{e,\mathrm{sep}}$ to the divertor neutral pressure have been observed experimentally also in TCV~\cite{Stagni_2022}, JET and Alcator C-Mod (C-Mod)~\cite{Silvagni_2025}, and in STEP SOLPS simulations~\cite{Henderson_2025}. 

It is important to note that $p_{0,\mathrm{div}}$ can be regarded as an engineering parameter, since it is primarily set by the applied gas puff level and effective pumping speed of the system~\cite{Kallenbach_2018, Moscheni_2026}, allowing for $n_{e,\mathrm{sep}}$ predictions based only on engineering parameters. Indeed, in AUG this approach enabled core-pedestal-SOL integrated modeling of plasma discharges using solely engineering parameters as input~\cite{Luda_2020}. However, the $n_{e,\mathrm{sep}}$ formulas used so far always depend on tokamak-specific constants, which make extrapolation efforts challenging. Also, single-machine studies cannot unveil the size dependence of $n_{e,\mathrm{sep}}$, an important topic for extrapolation.

To fill this gap, a multi-machine database of H-mode $n_{e,\mathrm{sep}}$ values evaluated with the same procedure has been assembled, using data from three metal-wall devices: Alcator C-Mod, ASDEX Upgrade and JET. The collected data have been used to derive a cross-machine scaling of $n_{e,\mathrm{sep}}$, with machine specific multiplication constants. Regression analysis revealed that $n_{e,\mathrm{sep}}$ reduces with increasing plasma minor radius (i.e. machine size), consistent with the associated increase of SOL connection length. The dependence of $n_{e,\mathrm{sep}}$ to other engineering parameters, such as $p_{0, \mathrm{div}}$, $I_p$, $B_t$, $P_{\mathrm{SOL}}/R$, is also discussed and compared to the $n_{e,\mathrm{sep}}$ expression given by the two-point model. Overall, the agreement found is remarkable, despite slight discrepancies in the exponents of $p_{0, \mathrm{div}}$ and $B_t$ which are further discussed. Perhaps more importantly, the two-point model multiplicative constant in the $n_{e,\mathrm{sep}}$ expression is compared to the regression-inferred tokamak-specific constants, finding good agreement. This allowed the introduction of a predictive $n_{e,\mathrm{sep}}$ formula based on the two-point model multiplicative constant and regression-based dependencies which proved able to predict $n_{e,\mathrm{sep}}$ within a factor 1.5 across the three devices. Predictions to next-step devices such as ITER, SPARC, DTT, JT-60SA and COMPASS-U are carried out, finding $n_{e,\mathrm{sep}}$ values in agreement with available SOLPS simulations. 

The content is organized as follows. In section~\ref{sec:db}, the database is described and correlations between parameters are discussed. In section~\ref{sec:reg}, the cross-machine $n_{e,\mathrm{sep}}$ scaling is introduced. In section~\ref{sec:2pt}, the two-point model expression for $n_{e,\mathrm{sep}}$ is derived and compared with the regression findings. The predictive formula and its applicability are discussed in section~\ref{sec:form}, while predictions to next-step devices are presented in section~\ref{sec:pred}. The conclusions are summarized in section~\ref{sec:concl}.

\section{Database}
\label{sec:db}
The database used in this work builds on the one assembled in~\cite{Silvagni_2025}, with some additional features that are discussed at the end of this section. The key characteristics of this database are:
\begin{enumerate}
    \item Use of similar closed divertor (lower single null) configurations across the three devices; 
    \item H-mode plasmas in favorable configuration---in JET with ELMs, in C-Mod without ELMs (the so-called EDA H-mode~\cite{Greenwald_1999}), while in AUG a mixture of ELMy and no-ELM plasmas;
    \item Unseeded plasmas in JET and C-Mod, while both unseeded and seeded plasmas in AUG;
    \item Stationary phases in both plasma and sub-divertor neutral pressure, the latter measured by baratrons along vertical pipes, see Fig.\,1 in Ref.~\cite{Silvagni_2025};
    \item Separatrix parameters evaluated with exactly the same procedure based on power balance in each device.
\end{enumerate}

\begin{table}
\centering
\footnotesize
\begin{tabular}{l|c|c|c}
    \hline\hline
     & C-Mod & AUG & JET \\
    \hline
    Discharges & 111 & 43 & 45  \\
    $I_{\mathrm{p}}$ (MA) & 0.6--1.4 & 0.6--1.2 & 1.8--2.6  \\
    $B_{\mathrm{t}}$ (T) & 4.5--7.8 & 1.8--2.6 & 2.1--3.3 \\
    $q_{95}$ & 2.9--7.2 & 3.1--6.9 & 2.9--4.0 \\
    $P_{\mathrm{SOL}}$ (MW) & 0.3--2.5 & 0.7--12.8 & 4.9--18.8 \\
    %$P_{\mathrm{SOL}} / R_{\mathrm{geo}}$ (MW/m) & 0.4--3.7 & 0.4--7.8 & 1.7--6.4 \\
    $\overline{n}_e$ ($10^{19}$ m$^{-3}$) & 16.9--53.6 & 4.8--13.2 & 3.9--8.5  \\
    $p_{0,\mathrm{div}}$ (Pa) & 0.7--16.9 & 0.2--7.2 & 0.03--0.9  \\
    %$\delta_{\mathrm{av}}$ & 0.4--0.5 & 0.2--0.4 & 0.2--0.3 \\
    $R_{\mathrm{geo}}$ (m) & 0.67--0.68 & 1.61--1.65 & 2.89--2.94  \\
    $a_{\mathrm{geo}}$ (m) & 0.21--0.22 & 0.49--0.53 & 0.89--0.95  \\
    \hline\hline
\end{tabular}
\caption{Parameter range of the assembled database.}
\label{table:db}
\end{table}

\noindent Concerning the last point, edge electron temperature and density profiles measured by Thomson scattering have been mapped to the outer midplane and fit with the same function in each device. Then, scrape-off layer (SOL) power balance has been applied to determine the electron temperature at the separatrix $T_{e,\mathrm{sep}}$, ensuring that the same pre-factors are used when evaluating $T_{e,\mathrm{sep}}$. 
A notable aspect of this power balance analysis is the direct evaluation of the inter-ELM temporal derivative of the plasma stored energy, $d\mathrm{W}/dt$, via fast magnetic equilibrium reconstruction---a term that is often neglected or approximated as a constant fraction of the absorbed heating power. For additional details on the separatrix evaluation procedure, the reader is referred to~\cite{Silvagni_2025}.
\newline The main plasma parameters of the assembled database are summarized in Table~\ref{table:db}. Compared to the database shown in~\cite{Silvagni_2025}, 
\begin{figure*}[hbt]
        \centerline{\includegraphics[width=0.9 \textwidth]{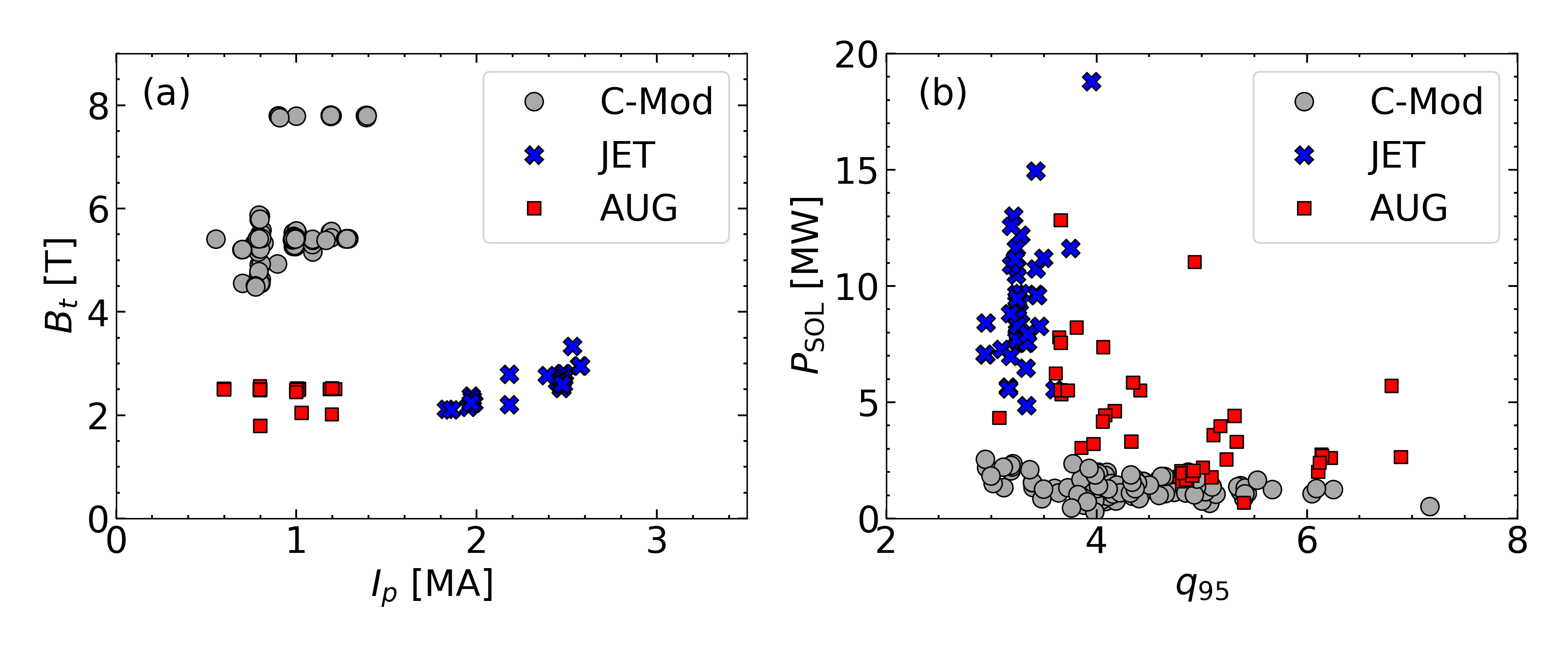}}
        \caption[]{On-axis toroidal magnetic field vs. plasma current (a) and power crossing the separatrix vs. edge safety factor (b) for the analyzed C-Mod (circles), AUG (squares) and JET (crosses) datasets.}
        \label{database}
\end{figure*} 
additional high-field plasmas at $B_t \approx 7.8 $\,T from C-Mod have been included~\cite{Tolman_2018}. Further, the JET dataset has been restricted to low triangularity plasmas only ($\delta_{\mathrm{av}} < 0.26$). The reason for this choice resides in the additional dependency of $n_{e,\mathrm{sep}}$ on triangularity found in JET `vertical-vertical' (V-V) target configuration~\cite{Silvagni_2025}, which is the divertor configuration used in this study for JET. This dependency deserves a detailed study that goes beyond the scope of this paper, therefore it is left as a topic for future research.
\newline Figure~\ref{database} shows the achieved variation in plasma current ($I_p$), on-axis toroidal magnetic field ($B_t$), safety factor evaluated at the 95\,$\%$ magnetic flux surface ($q_{95}$) and power entering the SOL ($P_{\mathrm{SOL}}$) in the three devices.
A satisfactory variation and de-correlation of $I_p$ and $B_t$ is obtained in AUG and C-Mod, which is reflected in the broad range of $q_{95}$ values spanned in both devices. In JET, while a good variation in $I_p$ and $B_t$ is obtained, these two values are mainly correlated, resulting in a small variation of $q_{95}$. Historically, this is due to the fact that experiments in V-V configuration at JET were mainly executed to be self-similar to ITER~\cite{Giroud_2015, Giroud_2024}, which is expected to operate a low $q_{95}$. The power entering
\begin{figure}
        \centerline{\includegraphics[width=0.5 \textwidth]{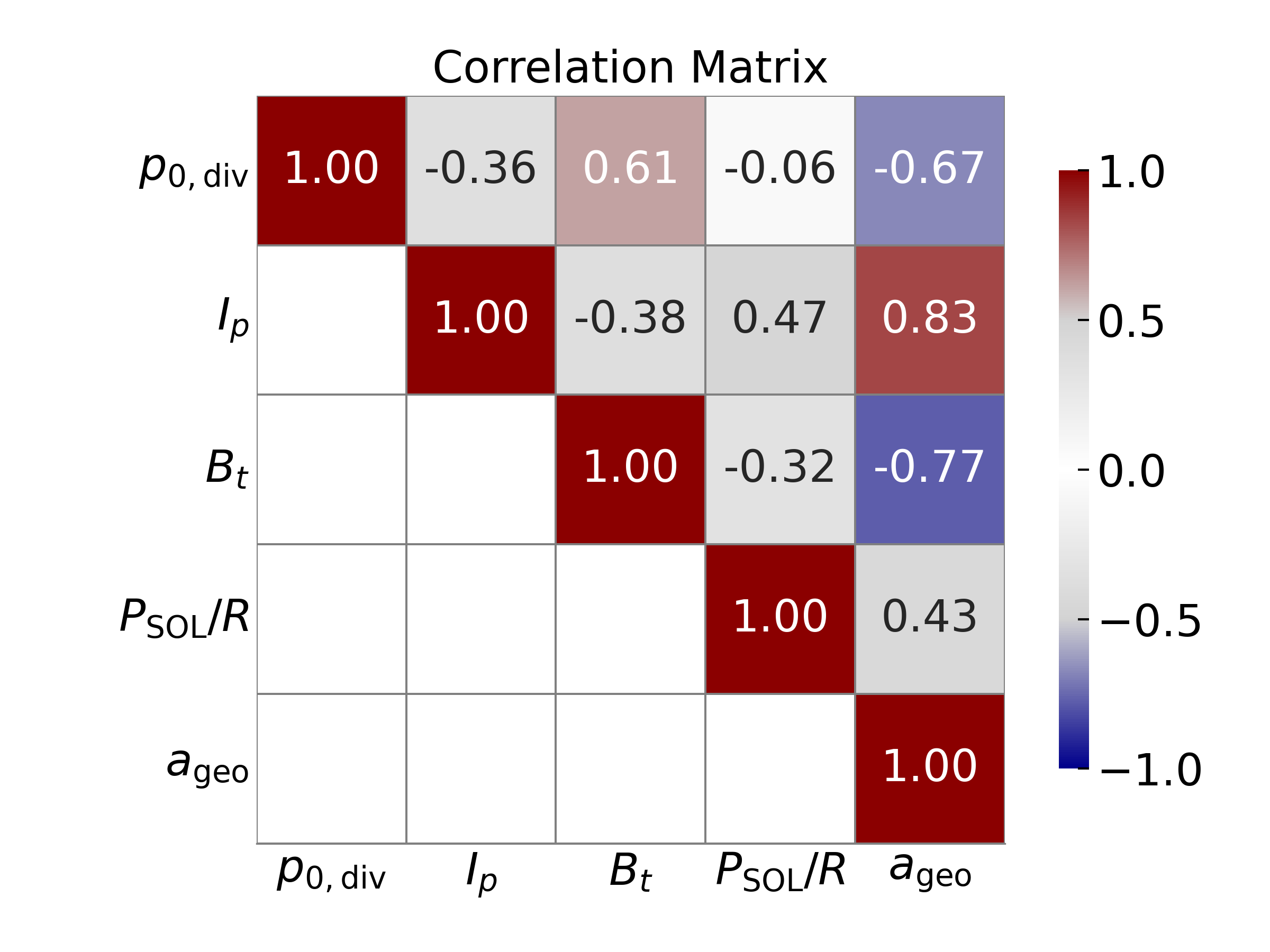}}
        \caption[]{Correlation matrix of the variables used for the regression analysis.}
        \label{corr_matrix}
\end{figure} 
the SOL exhibits also a good variation, as well as the line-averaged density ($\overline{n}_{e}$) which is spanning more than one order of magnitude across the three devices, see table~\ref{table:db}. The divertor neutral pressure $p_{0,\mathrm{div}}$, measured by a baratron located in the sub-divertor region~\cite{Silvagni_2025}, varies by about two orders of magnitude. As it is also shown in Appendix A, the divertor neutral pressure $p_{0,\mathrm{div}}$ can be regarded as an engineering parameter set primarily by the injected particle flux and the effective pumping speed of the system~\cite{Kallenbach_2018, Moscheni_2026}.
\newline The tokamaks analyzed in this study have a nearly constant aspect ratio $A = R_{\mathrm{geo}}/a_{\mathrm{geo}} = 3.0-3.3$, where $R_{\mathrm{geo}}$ and $a_{\mathrm{geo}}$ are the geometrical major and minor radii, respectively. As a result, $R_{\mathrm{geo}}$ and $a_{\mathrm{geo}}$ are fully correlated, and cannot be used together as independent variables in the regression analysis. Therefore, we use only $a_{\mathrm{geo}}$ as a regression variable in this study, while $R_{\mathrm{geo}}$ is employed solely as a normalization factor for $P_{\mathrm{SOL}}$. This choice is motivated by theoretical considerations, which will be discussed further in section~\ref{sec:2pt}.
\newline Figure~\ref{corr_matrix} shows the correlation matrix of the variables chosen for regression analysis. A good level of de-correlation is achieved between $I_p$, $B_t$, $P_{\mathrm{SOL}}$ and $p_{0,\mathrm{div}}$. The same cannot be said for $a_{\mathrm{geo}}$, which shows a positive correlation with $I_p$ and a negative correlation with $B_t$. Therefore, particular care should be taken with the $a_{\mathrm{geo}}$ dependencies found in this study.

\section{Multi-machine scaling of the separatrix electron density}
\label{sec:reg}
To model the variable $n_{e,{\mathrm{sep}}}$, the following power law is considered $n_{e,\mathrm{sep}} = C_{\mathrm{dev}} \cdot p_{0, \mathrm{div}}^{\alpha_{p}} \cdot I_{p}^{\alpha_I} \cdot B_{t}^{\alpha_B} \cdot \left( \frac{P_{\mathrm{SOL}}}{R_{\mathrm{geo}}}\right)^{\alpha_{P/R}} \cdot a_{\mathrm{geo}}^{\alpha_a} $, where $C_{\mathrm{dev}}$ is a device-specific multiplication constant and $\alpha_{X}$ is the exponent of the independent variable $X$. This expression has been chosen to facilitate comparisons with the two-point model, as will be discussed in the next section.
Applying a generalized linear model with Gaussian likelihood and the logarithmic link function, the following scaling expression is obtained:
\begin{gather}
    \nonumber n_{e,\mathrm{sep,sc}} [10^{19} \mathrm{m}^{-3}] = C_{\mathrm{dev}} \cdot (p_{0, \mathrm{div}} [\mathrm{Pa}])^{0.20 \pm 0.03} \cdot (I_{p} [\mathrm{MA}])^{0.03 \pm 0.09} \cdot (B_{t} [\mathrm{T}])^{-0.26 \pm 0.11} \cdot \\ 
    \left( \frac{P_{\mathrm{SOL}}}{R_{\mathrm{geo}}} \left[\frac{\mathrm{MW}}{\mathrm{m}}\right]\right)^{0.19 \pm 0.04} \cdot (a_{\mathrm{geo}} [\mathrm{m}])^{-0.47 \pm 1.92},
    \label{eq:nsepsc}
\end{gather}
\begin{figure}
        \centerline{\includegraphics[width=0.5 \textwidth]{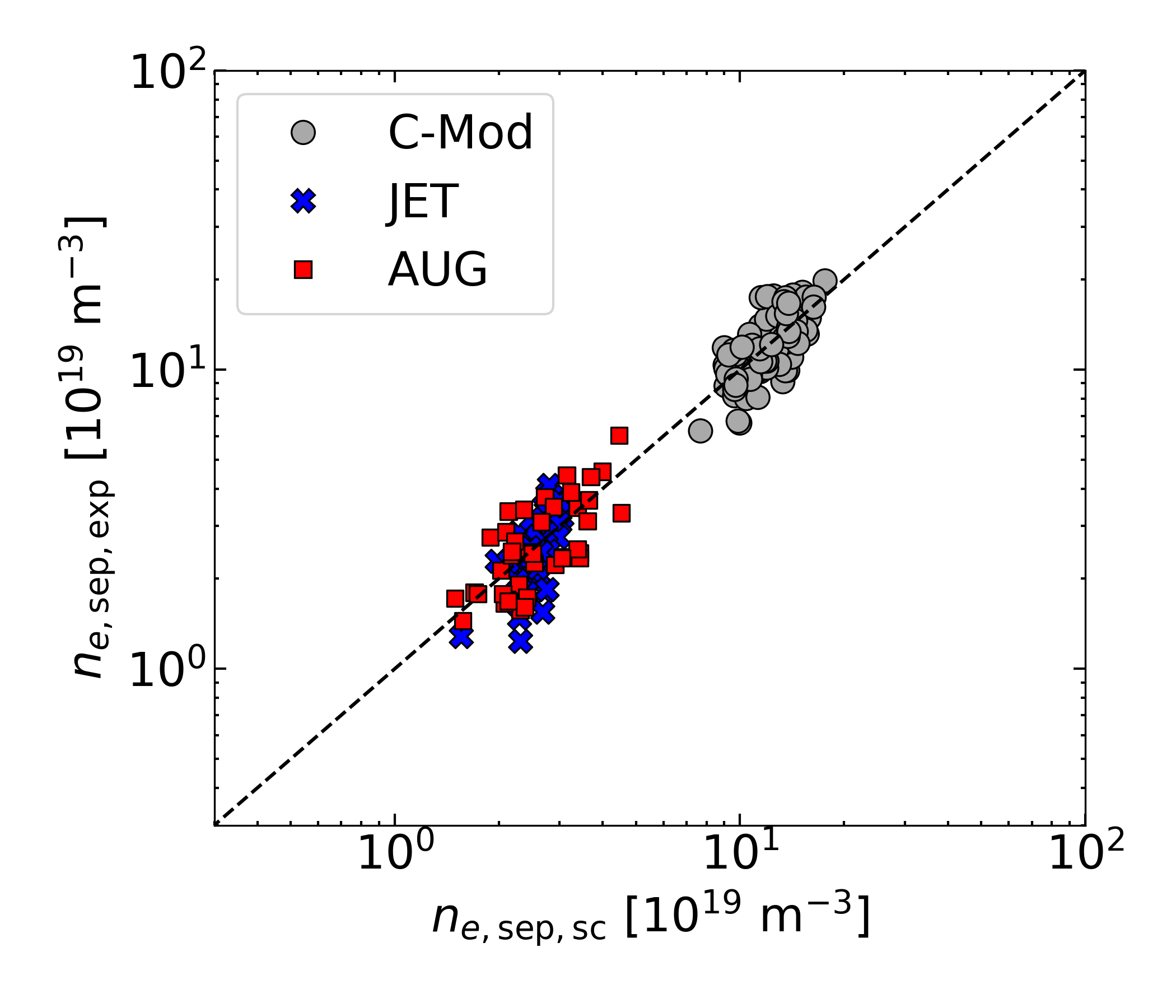}}
        \caption[]{Experimental $n_{e,\mathrm{sep}}$ values against those predicted by Eq.~\ref{eq:nsepsc} for C-Mod (gray circles), AUG (red squares) and JET (blue crosses).}
        \label{nsep_scaling}
\end{figure} 
with $C_{\mathrm{dev}}$ being $6.3 \pm 2.4$ for C-Mod, $ 2.0 \pm 0.9$ for AUG, and $ 3.0 \pm 2.8$ for JET. The normalized root mean square error is NRMSE = 19\,$\%$, while $R^2= 0.91$.
Figure~\ref{nsep_scaling} shows the comparison between the experimental $n_{e,\mathrm{sep}}$ data and those predicted by the scaling. A satisfactory agreement is found across more than one order of magnitude. Regression analysis reveals that the two parameters driving an increase in $n_{e,\mathrm{sep}}$ are the divertor neutral pressure and $P_{\mathrm{SOL}}/R_{\mathrm{geo}}$, both exhibiting similar exponents ($\sim 0.2$). 
The observed dependence of $n_{e,\mathrm{sep}}$ on $P_{\mathrm{SOL}}/R_{\mathrm{geo}}$ aligns with previous studies at JET~\cite{Lomanowski_2023, Erents_2000_2} and DIII-D~\cite{Leonard_2017}, which also reported a mild positive correlation between the two quantities.
\newline The dependence on $p_{0,\mathrm{div}}$ is consistent with results from single-machine analyses~\cite{Kallenbach_2018, Silvagni_2025}, with an exponent that most closely resembles the value reported for C-Mod. 
\begin{figure*}[hbt]
        \centerline{\includegraphics[width=1 \textwidth]{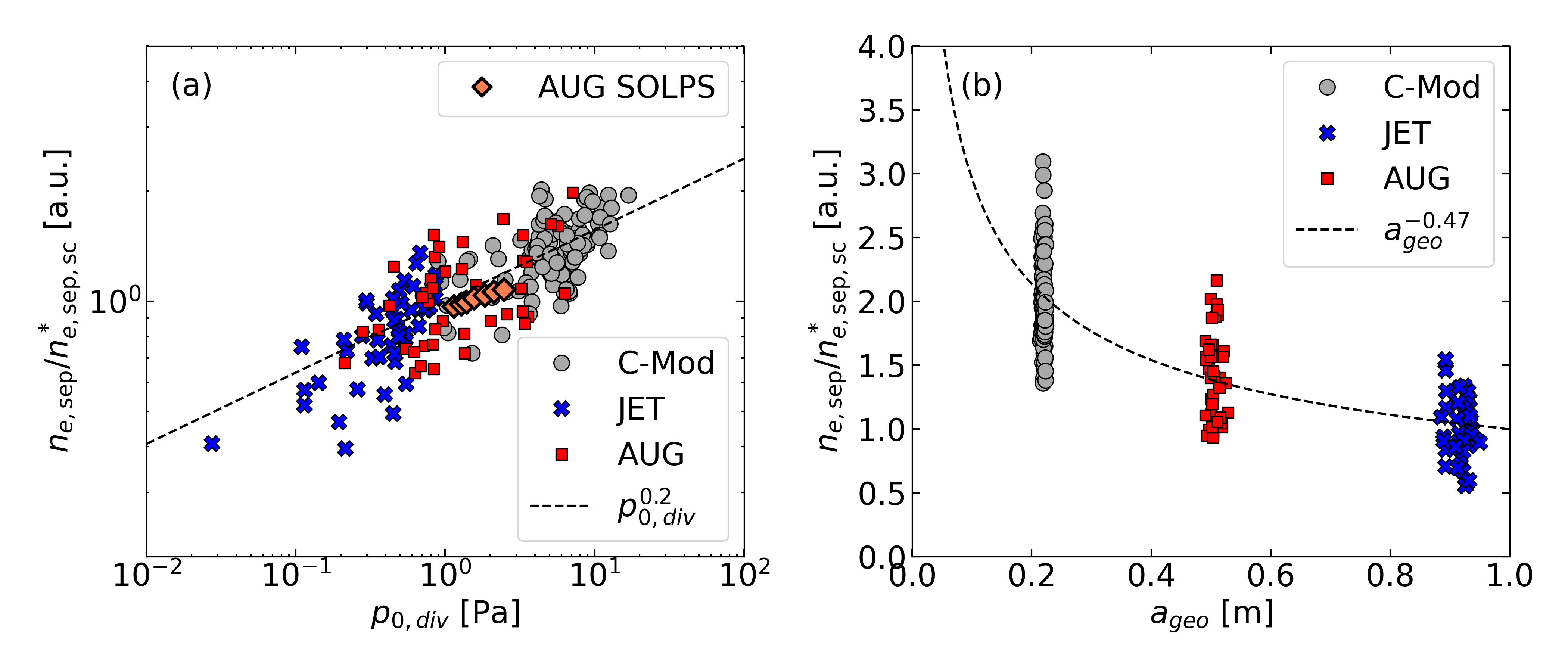}}
        \caption[]{Experimental $n_{e,\mathrm{sep}}$ values normalized by the scaling expression in Eq.~\ref{eq:nsepsc}, excluding the dependence on the parameter shown on the x-axis: $p_{0,\mathrm{div}}$ in (a) and $a_{\mathrm{geo}}$ in (b). The residual scaling dependence on $p_{0,\mathrm{div}}$ and $a_{\mathrm{geo}}$ is plotted as a dashed line. Results from SOLPS gas scan simulations of an AUG H-mode plasma are shown in light red diamonds in panel (a).}
        \label{nsepnorm}
\end{figure*} 
Moreover, AUG SOLPS gas scan simulations confirm the experimental relationship found between $n_{e,\mathrm{sep}}$ and $p_{0,\mathrm{div}}$, see Fig.~\ref{nsepnorm}\,(a). The simulated discharge is a typical H-mode AUG plasma belonging to the database ($I_p = 1 $\,MA, $B_t = 2.5$\,T, $P_{\mathrm{SOL}} = 6$\,MW), where particle and heat diffusion coefficients have been chosen to fit upstream and downstream profiles. The subdivertor structure is similar to the one employed in Ref.~\cite{Zito_2025}, yielding a good match with the experimental subdivertor neutral pressure. Then, the input gas puff was scanned keeping constant transport coefficients. The obtained relationship between the output separatrix densities and neutral pressures is in good agreement with the experimental results, similarly to previous findings with SOLPS simulations~\cite{Kallenbach_2019}.
Furthermore, a comparable relationship ($n_{e,\mathrm{sep}} \propto p_{0,\mathrm{div}}^{0.22}$) has also been found in recent SOLPS-ITER gas scan simulations of the ITER Q=10 baseline scenario~\cite{Pshenov_2025}, increasing confidence in the applicability of this scaling in plasmas with high neutral opacity. 
\newline The two parameters that lead to a decrease in $n_{e,\mathrm{sep}}$ when increased are the plasma minor radius and the toroidal magnetic field $B_t$. A negative dependence on $B_t$ was also observed experimentally in DIII-D~\cite{Leonard_2017}, and was correlated with an increase of the SOL radiative fraction as $B_t$ increases. Additionally, several tokamaks have reported a reduction in $n_{e,\mathrm{ped}}$ with increasing $B_t$ at fixed plasma current~\cite{Hughes_2002, Tolman_2018, Leonard_2017, Frassinetti_2025}, a trend that would be consistent with a corresponding decrease in $n_{e,\mathrm{sep}}$. 
\newline The other quantity exhibiting a relatively strong negative correlation with $n_{e,\mathrm{sep}}$ is $a_{\mathrm{geo}}$, as also shown in Fig.~\ref{nsepnorm}\,(b). This finding will be compared to the two-point model expectations in the next section. However, as expected from the discussion in section~\ref{sec:db}, a large error is associated to the $a_{\mathrm{geo}}$ exponent, which calls for particular caution in the interpretation of this dependency.
\newline Lastly, the multi-machine regression finds no relevant dependence of $n_{e,\mathrm{sep}}$ on the plasma current $I_p$. This result is consistent with previous observations in each single device~\cite{Kallenbach_2018, Silvagni_2025}, which also reported an absent correlation with $I_p$ upon normalizing $n_{e,\mathrm{sep}}$ to the $p_{0,\mathrm{div}}$ dependency. It should be noted that when $n_{e,\mathrm{sep}}$ is not normalized to the $p_{0,\mathrm{div}}$ dependency, a mild correlation with the plasma current is typically observed~\cite{Kallenbach_2018, Leonard_2017, Harrison_2024}. However, this mild correlation is thought to be due to the larger gas puff usually required at higher current to avoid impurity accumulation in H-mode. Indeed, employing a Z-test to our variables reveals that $p_{0,\mathrm{div}}$ is a very strong statistical predictor (p-value of zero), while $I_p$ is not a statistically significant predictor (p-value of 0.735).

\section{Two-point model expression for $n_{e,\mathrm{sep}}$ as a function of engineering parameters}
\label{sec:2pt}

In this section, an expression for $n_{e,\mathrm{sep}}$ as a function of engineering parameters will be derived. This treatment is an extension of the one introduced by Kallenbach et al. in Ref.~\cite{Kallenbach_2018}. A SOL flux tube connecting the outer midplane to the outer divertor target is considered, as typically done with two-point modeling~\cite{Stangeby_2018}. Key assumptions of the model used here are: (1) Ions and electrons have \DEdit{the same}\Edit{distinct} temperature\Edit{s and densities}, \Edit{$T_i$, $T_e$, $n_i$, $n_e$}\DEdit{$T_i = T_e = T$}, with temperatures expressed in eV; (2) The upstream parallel velocity is zero, while the velocity at the target is the ion sound speed \Edit{defined as $c_s = \sqrt{\frac{(1+\tau / z)e T_e}{m_i}}$~\cite{Stangeby_2018}}\DEdit{for an isothermal flow $c_s = \sqrt{\frac{2e T}{m_i}}$}, where $e$ is the electron charge\Edit{,}\DEdit{and} $m_i$ the ion mass\Edit{, $\tau=T_i/T_e$ and $z = n_e/n_i$}; (3) All the power enters the flux tube at the upstream position and is transported in the parallel direction solely by electron conduction. Momentum and power sinks along the flux tube are parametrized by loss factors, defined as $(1-f_{\mathrm{mom}}) p_{\mathrm{tot,u}} = p_{\mathrm{tot,d}}$ and $(1-f_{\mathrm{pow}}) P_{\mathrm{u}} = P_{\mathrm{d}}$ for the momentum and power, respectively. The subscripts `u' and `d' stand for `upstream' and `downstream' positions along the flux tube, while $P$ indicates the power and $p_{\mathrm{tot}}$ the total plasma pressure, given by the sum of static and dynamic pressures.
\newline The four basic equations of the two-point model are given by momentum and power conservation, Fourier's law to describe heat transport in the parallel direction, and a target boundary condition, which defines the target heat flux at the sheath entrance. These equations read:
\begin{alignat}{2}
    (1-f_{\mathrm{mom}}) \Edit{(1+\tau_{\mathrm{u}} / z_{\mathrm{u}})} n_{\mathrm{u}} T_{\mathrm{u}} &= 2 \Edit{(1+\tau_{\mathrm{d}} / z_{\mathrm{d}})} n_{\mathrm{d}} T_{\mathrm{d}} \label{eq:mom} \\
    (1-f_{\mathrm{pow}}) q_{||,\mathrm{u}} A_{\perp, \mathrm{SOL,u}} &= q_{||,\mathrm{d}} A_{\perp, \mathrm{SOL,d}} \quad \Rightarrow \quad (1-f_{\mathrm{pow}}) q_{||,\mathrm{u}} = q_{||,\mathrm{d}} b \frac{B_{\mathrm{u}}}{B_{\mathrm{d}}} \label{eq:pow} \\
    T_{u} &= \left( \frac{7 q_{||,\mathrm{u}} L_{||}}{2 \kappa_e} \right)^{2/7}  \label{eq:fourier} \\
    q_{||,\mathrm{d}} &= \gamma e T_{\mathrm{d}} \Gamma_{||,\mathrm{d}} \quad \Rightarrow \quad q_{||,\mathrm{d}} = \gamma e T_{\mathrm{d}} n_{\mathrm{d}}  c_{s, \mathrm{d}}\label{eq:sheath}
\end{alignat}
where $q_{||}$ indicates the heat flux parallel to the magnetic field, $A_{\perp, \mathrm{SOL}}$ the SOL flux bundle area perpendicular to the magnetic field line, $B$ the total magnetic field, $L_{||}$ the SOL parallel connection length between the outer midplane and the target, $\kappa_e$ is the parallel electron conductivity coefficient, $\gamma$ is the total sheath heat transmission factor, $\Gamma_{||,\mathrm{d}} = n_{\mathrm{d}}  c_{s, \mathrm{d}}$ is the parallel ion flux density (particles/m$^2$/s) at the sheath entrance, and $b = \lambda_{\mathrm{int}}/ \lambda_{q} = 1 + 1.64 S / \lambda_q$ is the divertor broadening factor, with $\lambda_{q}$ and $\lambda_{\mathrm{int}}$ being the outer midplane heat flux decay length and the integral heat flux decay length at the target, respectively, while $S$ is the broadening parameter. For a detailed overview of $\lambda_{q}$, $\lambda_{\mathrm{int}}$ and $S$, the reader is referred to Ref.~\cite{Eich_PRL2011, Eich_2013, makowski2012analysis}. Compared to the equations typically used for the two-point model~\cite{Stangeby_2018, Kallenbach_2018,  Lomanowski_2023}, the main difference is in the factor $B_{\mathrm{u}} / B_{\mathrm{d}}$ in Eq.~\ref{eq:pow}, which is typically approximated to be $ B_{t,\mathrm{u}} / B_{t,\mathrm{d}} \sim R_{\mathrm{d}} / R_{\mathrm{u}}$. However, to potentially model also alternative divertor configurations~\cite{Lunt_2017, Lunt_2023} which could have different $B_{p, \mathrm{u}} / B_{p, \mathrm{d}}$ ratios, this approximation is not made here. Also, it is worth noting that the broadening factor $b$ accounts for power losses perpendicular to the magnetic flux bundle in the divertor region. Therefore, in this formulation, the power loss factor $f_{\mathrm{pow}}$ primarily reflects losses due to radiation and charge exchange reactions. A full derivation of Eq.~\ref{eq:pow} is described in the Appendix B. 
\newline Combining Eq.~\ref{eq:mom}, ~\ref{eq:pow}, ~\ref{eq:fourier} and~\ref{eq:sheath}, an expression for the upstream electron density $n_{e,\mathrm{u}}$ as a function of $q_{||,\mathrm{u}}$, $L_{||}$ and $\Gamma_{||,\mathrm{d}}$ is obtained:
\begin{equation}
    n_{e,\mathrm{u}} = \frac{2 (1-f_{\mathrm{pow}})^{1/2}}{(1-f_{\mathrm{mom}})} \Edit{\frac{(1+\tau_{\mathrm{d}} /z_{\mathrm{d}})^{1/2}}{(1+\tau_{\mathrm{u}} /z_{\mathrm{u}})}} \left(\frac{2 \kappa_e}{7}\right)^{2/7} \frac{1}{e} \left( \frac{B_{\mathrm{d}}}{B_{\mathrm{u}}}   \frac{m_i}{\gamma \DEdit{2}  b}\right)^{1/2} q_{||,\mathrm{u}}^{3/14} L_{||}^{-2/7} \Gamma_{||,d}^{1/2},
    \label{eq:nsep_phys}
\end{equation}
which differs from the one derived in~\cite{Kallenbach_2018} \DEdit{only} by the additional $\sqrt{B_{\mathrm{d}}/B_{\mathrm{u}}}$ \Edit{and $1+\tau / z$} factor\Edit{s}.
This expression highlights the primary physics parameters affecting $n_{e,\mathrm{sep}}$. Both the ion flux density at the target and the upstream heat flux drive an increase in $n_{e,\mathrm{sep}}$, while extending the connection length between the outer midplane and the target reduces $n_{e,\mathrm{sep}}$. While Eq.~\ref{eq:nsep_phys} is extremely insightful to study what drives changes in $n_{e,\mathrm{sep}}$, it \Edit{is not directly comparable with the regression, as the dependencies on engineering parameters are not made explicit.}\DEdit{is of reduced usage for what concerns extrapolations, as the ion flux density at the target can be estimated only with the help of modeling.} Therefore, in the next paragraphs a series of assumptions will be introduced to express $n_{e,\mathrm{sep}}$ as a function of engineering parameters.
\newline To express $q_{||,\mathrm{u}}$ and $L_{||}$ as a function of $I_p$ and $B_t$ (on-axis toroidal magnetic field), the poloidally-averaged poloidal magnetic field $\langle B_p \rangle = \mu_0 I_p / (2 \pi a_{\mathrm{geo}} \hat{\kappa}) $ and the cylindrical safety factor 
\begin{equation}
    q_{\mathrm{cyl}} = \frac{\hat{\kappa} a_{\mathrm{geo}} B_t } {R_{\mathrm{geo}} \langle B_p \rangle} = \frac{2 \pi}{ \mu_0} \frac{\hat{\kappa}^2 a^2_{\mathrm{geo}} B_t }{R_{\mathrm{geo}} I_p}
    \label{eq:qcyl}
\end{equation}
are introduced, with 
\begin{equation}
    \hat{\kappa} = \sqrt{\frac{1+\kappa^2 (1+ 2 \delta^2 - 1.2 \delta^3) }{2}}
    \label{eq:kappahat}
\end{equation}
being an effective plasma elongation corrected for the triangularity, a parameter introduced in Ref.~\cite{Uckan_1990, Uckan_1991} to account for the non-circular plasma cross-section.
Alternative definitions of the cylindrical safety factor have been adopted in the literature and are discussed in Appendix C. 
Using the above-mentioned formulas, $L_{||}$ can be expressed as:
\begin{equation}
    L_{||} = l^* \pi R_{\mathrm{geo}} q_{\mathrm{cyl}} = l^* \frac{2}{\mu_0} (\pi a_{\mathrm{geo}} \hat{\kappa})^2 \frac{B_t}{I_p},
    \label{eq:Lsol}
\end{equation}
where $l^*$ is a constant describing the deviation of the real connection length from the approximated formula $\pi R_{\mathrm{geo}} q_{\mathrm{cyl}}$. For the divertor configurations analyzed in this study, $l^*$ is about 1.2, 1.3, and 1.1 for C-Mod, AUG and JET, respectively. To calculate this number the SOL connection length has been evaluated from the magnetic field line at approximately half $\lambda_q$~\cite{Silvagni_2025}. Clearly, $l^*$ can be larger when alternative divertor configurations are considered.  
\newline The upstream parallel heat flux can be rewritten as:
\begin{equation}
    q_{||,\mathrm{u}} = \frac{f_{\mathrm{out}} P_{\mathrm{SOL}}}{A_{\perp, \mathrm{SOL, u}}} = \frac{f_{\mathrm{out}} P_{\mathrm{SOL}} B_{\mathrm{u}}}{2 \pi R_{\mathrm{u}} \lambda_q B_{p, \mathrm{u}}} = \frac{f_{\mathrm{out}} P_{\mathrm{SOL}} B_{t}}{C_{\mathrm{geo}} 2 \pi R_{\mathrm{geo}} \lambda_q \langle B_{p} \rangle} =\frac{f_{\mathrm{out}}  P_{\mathrm{SOL}} \hat{\kappa} a_{\mathrm{geo}}  B_{t} }{\mu_0 C_{\mathrm{geo}} \lambda_q  R_{\mathrm{geo}}  I_p},
    \label{eq:qu}
\end{equation}
where $f_{\mathrm{out}}$ is the fraction of $P_{\mathrm{SOL}}$ going towards the outer target, and $C_{\mathrm{geo}}$ is a geometrical constant describing the ratio of quantities between outer midplane and geometrical axis: 
\begin{equation}
     C_{\mathrm{geo}} = \frac{R_{\mathrm{u}}}{R_{\mathrm{geo}}} \frac{B_{t}}{B_{\mathrm{u}}} \frac{B_{p,\mathrm{u}}}{\langle B_{p} \rangle} \approx \frac{(R_{\mathrm{geo}}+a_{\mathrm{geo}})^2}{R_{\mathrm{geo}}^2}  \frac{B_{p,\mathrm{u}}}{\langle B_{p} \rangle} , 
    \label{eq:Cgeo}
\end{equation}
where, in the last expression, $B_{\mathrm{u}} \approx B_{t,\mathrm{u}}$ has been considered. In this work, the exact formula of $C_{\mathrm{geo}}$ is used, which gives values ranging between 2.3 and 2.5 for the three tokamaks.
\newline Following the same argument introduced in~\cite{Kallenbach_2018, Maingi_1999}, it is assumed that the neutral flux density, $\Gamma_0$, measured in the sub-divertor region equals the ion flux density perpendicular to the target, $\Gamma_{\perp, d}$. Assuming a Maxwellian distribution function of the neutrals at room temperature \Edit{and that neutrals are fully recombined into molecules at this temperature}, the \Edit{molecular} neutral flux density can be converted to a neutral pressure via the formula $\Gamma_0 = \Edit{2} n_0 \frac{1}{4} \sqrt{\frac{8 e T_0} {\pi m_0 }} = \Edit{\sqrt{\frac{2}{\pi m_0 e T_0} }} p_0 = C_0 p_0$, where $C_0 = \Edit{\sqrt{\frac{2} {\pi m_0 e T_0}}  = 1.51} \times 10^{23} $ atoms/m$^2$/s/Pa for deuterium\Edit{,}\DEdit{and} $T_0 = 300$\,K \Edit{and $m_0 = 2m_i$ is the molecule mass}. Therefore, the parallel ion flux density can be rewritten as:
\begin{equation}
    \Gamma_{||,d} = \frac{\Gamma_{\perp,d}}{\mathrm{sin}(\alpha_{\mathrm{div}})} \approx \frac{\Gamma_{0}}{\mathrm{sin}(\alpha_{\mathrm{div}})} = \frac{C_{0} p_0}{\mathrm{sin}(\alpha_{\mathrm{div}})},
    \label{eq:Gamma}
\end{equation}
where $\alpha_{\mathrm{div}}$ is the magnetic field line grazing angle at the outer target. Inserting Eq.~\ref{eq:Lsol}, ~\ref{eq:qu} and~\ref{eq:Gamma} into Eq.~\ref{eq:nsep_phys}, the following expression for $n_{e,\mathrm{sep}}$ as a function of engineering parameters is obtained:
\begin{equation}
    n_{e,\mathrm{sep}} = C_{\mathrm{2pt}} \left( \frac{P_{\mathrm{SOL}}}{R_{\mathrm{geo}}}\right)^{3/14} I_p^{1/14} B_t^{-1/14} a_{\mathrm{geo}}^{-5/14} p_{0,\mathrm{div}}^{1/2},
    \label{eq:nsep_eng}
\end{equation}
with $C_{\mathrm{2pt}}$ being:
\Edit{\begin{gather}
    \nonumber C_{\mathrm{2pt}} = \frac{2 (1-f_{\mathrm{pow}})^{1/2}}{(1-f_{\mathrm{mom}})} \frac{(1+\tau_{\mathrm{d}} /z_{\mathrm{d}})^{1/2}}{(1+\tau_{\mathrm{u}} /z_{\mathrm{u}})} \left(\frac{\kappa_e}{7 l^*}\right)^{2/7} \frac{\mu_0^{1/14} 10^{12/7}}{\pi^{4/7} e } \\ \left( \frac{f_{\mathrm{out}}}{C_{\mathrm{geo}} \lambda_q}\right)^{3/14} 
    \left( \frac{B_{\mathrm{d}}}{B_{\mathrm{u}}}   \frac{C_0 m_i}{ \gamma  b \cdot \mathrm{sin}(\alpha_{\mathrm{div}})}\right)^{1/2} \hat{\kappa}^{-5/14} .
    \label{eq:C2pt}
\end{gather}}
\begin{gather}
    \nonumber \DEdit{C_{\mathrm{2pt}} = 5.165 \times 10^{19}\frac{(1-f_{\mathrm{pow}})^{1/2}}{(1-f_{\mathrm{mom}})} \left(\frac{\kappa_e}{l^*}\right)^{2/7} \left( \frac{f_{\mathrm{out}}}{C_{\mathrm{geo}} \lambda_q}\right)^{3/14} \left( \frac{B_{\mathrm{d}}}{B_{\mathrm{u}}}   \frac{C_0 m_i}{ \gamma b  \cdot \mathrm{sin}(\alpha_{\mathrm{div}})}\right)^{1/2} \hat{\kappa}^{-5/14}.}
    %\label{eq:C2pt}
\end{gather}
In equations~\ref{eq:nsep_eng} and~\ref{eq:C2pt}, all quantities are expressed in SI units (hence $\lambda_q$ is expressed in meters), except for $I_p$ (MA) and $P_{\mathrm{SOL}}/R_{\mathrm{geo}}$ (MW/m)\Edit{, which is the reason for the factor $10^{12/7}$ in Eq.~\ref{eq:C2pt}}. 
\newline The exponents obtained for $P_{\mathrm{SOL}}/R_{\mathrm{geo}}$, $I_p$ and $a_{\mathrm{geo}}$ are similar to those found with regression analysis, which is remarkable given the simplicity of the two-point model. 
However, the exponents of $p_{0,\mathrm{div}}$ and $B_t$ are somewhat stronger and weaker in magnitude, respectively, than those found in the regression. This discrepancy could stem from additional hidden dependencies of the loss factors on these parameters. Indeed, in DIII-D it was found that the SOL radiative fraction increases as $B_t$ rises at constant current~\cite{Leonard_2017}. This behavior, which could be due to the additional SOL volume available due to the increased connection length, would reduce $(1-f_{\mathrm{pow}})$ as $B_t$ increases, potentially explaining the stronger negative $B_t$ exponent found in the regression. Similarly, SOL power losses are known to increase with decreasing plasma temperature in the divertor region $T_{\mathrm{d}}$~\cite{Stangeby_2018, Lomanowski_2023} or, equivalently, to increase as the neutral pressure rises~\cite{Osborne_2025}. This would cause $(1-f_{\mathrm{pow}})$ to decrease when $p_{0,\mathrm{div}}$ rises, which, in turn, could explain the smaller positive exponent of $p_{0,\mathrm{div}}$ found in the regression.
\newline On the other hand, the good agreement found between the two-point model and regression exponents for $P_{\mathrm{SOL}}/R_{\mathrm{geo}}$ and $a_{\mathrm{geo}}$ allows these dependencies to be interpreted more directly. The weak positive dependence on $P_{\mathrm{SOL}}/R_{\mathrm{geo}}$ can be attributed to the $n_{e,\mathrm{sep}}$ dependence on the upstream parallel heat flux $q_{||,\mathrm{u}}$ found in Eq.~\ref{eq:nsep_phys}. This relationship can be intuitively understood by considering that ionization is an energy-expensive process; a larger upstream heat flux helps sustain ionizations, thereby increasing the upstream density. Conversely, the negative dependence on $a_{\mathrm{geo}}$ found via regression can be mainly attributed to the SOL connection length dependence of $n_{e,\mathrm{sep}}$, as described by Eq.~\ref{eq:nsep_phys}. For fixed downstream plasma conditions, a longer $L_{||}$ increases the upstream temperature due to conduction (see Eq.~\ref{eq:fourier}) and, in turn, reduces $n_{e,\mathrm{sep}}$ to maintain the pressure balance. 
\newline Of particular interest for prediction capabilities is the comparison of the multiplicative constant obtained from the two-point model, $C_{\mathrm{2pt}}$, with the device-specific multiplicative constant inferred through regression, $C_\mathrm{dev}$. 
\Edit{First, Eq.~\ref{eq:C2pt} can be further simplified by introducing a series of additional assumptions. A low-Z impurity, such as boron, is assumed to be the main impurity in the plasma, and an effective ion charge $Z_{\mathrm{eff}} = 2$ is considered. Accordingly, $n_i/n_e = (6-Z_{\mathrm{eff}})/5 = 0.8$ and $z_{\mathrm{d}} = z_{\mathrm{u}} = 1.25$. Moreover, $\tau_{\mathrm{d}} = \tau_{\mathrm{u}} = 2$ is assumed, which is consistent with experimental measurements~\cite{Brunner_2013, Kocan_2008, Bucalossi_2025iaea}. Furthermore, the total sheath heat transmission factor $\gamma$ is fixed to 7~\cite{Stangeby_book, Brida_2020}. With these choices, Eq.~\ref{eq:C2pt} can be further simplified as follows:
\begin{gather}
    C_{\mathrm{2pt}} = 1.712 \times 10^{19}\frac{(1-f_{\mathrm{pow}})^{1/2}}{(1-f_{\mathrm{mom}})} \left(\frac{\kappa_e}{l^*}\right)^{2/7} \left( \frac{f_{\mathrm{out}}}{C_{\mathrm{geo}} \lambda_q}\right)^{3/14} \left( \frac{B_{\mathrm{d}}}{B_{\mathrm{u}}}   \frac{C_0 m_i}{ b  \cdot \mathrm{sin}(\alpha_{\mathrm{div}})}\right)^{1/2} \hat{\kappa}^{-5/14}.
    \label{eq:C2pt_simple}
\end{gather}
}
In Table~\ref{table:c2pt}, the main parameters used to evaluate $C_{\mathrm{2pt}}$ in the three machines are summarized. 
\begin{table}[htb]
\centering
\footnotesize
\begin{tabular}{l|c|c|c}
    \hline\hline
     & C-Mod & AUG & JET \\
    \hline
    $f_{\mathrm{out}}$ & 0.50 & 0.65 & 0.65  \\
    $l^*$ & 1.17 & 1.31 & 1.09  \\
    $\alpha_{\mathrm{div}}$ (degree) & 0.65 & 3.30 & 2.50 \\
    $S/\lambda_{q}$ & 1.00 & 0.75 & 0.50 \\
    $B_{\mathrm{d}} / B_{\mathrm{u}}$ & 1.42 & 1.32 & 1.29 \\
    $C_{\mathrm{geo}} $ & 2.34 & 2.51 & 2.29 \\
    $\hat{\kappa} $ & 1.48 & 1.40 & 1.43 \\
    \hline
    $C_{\mathrm{2pt}} (10^{19}) $ & \Edit{6.32}\DEdit{6.26} & \Edit{2.22}\DEdit{2.2} & \Edit{2.88}\DEdit{2.85} \\
    \hline\hline
\end{tabular}
\caption{Summary of parameter values used to evaluate $C_{\mathrm{2pt}}$ in the three tokamaks.}
\label{table:c2pt}
\end{table}
The quantities directly evaluated from the equilibrium are $l^*$, $\alpha_{\mathrm{div}}$, $B_{\mathrm{d}} / B_{\mathrm{u}}$, $C_{\mathrm{geo}} $ and $\hat{\kappa} $. In this study, the upstream position is defined as the outer midplane. The fraction of power flowing to the outer target, $f_{\mathrm{out}}$, has been chosen to be 0.5 in C-Mod~\cite{Terry_2013, Brunner_2018_powersharing}, and 0.65 in AUG and JET~\cite{Pitts_2005}, following experimental measurements. The values for $S/\lambda_q$, which in turn set the broadening factor $b = 1 + 1.64 S/\lambda_q$, are based on the experimental values found in the ITPA multi-machine database~\cite{Eich_2013}. The power fall-off length $\lambda_q$ has been estimated as $2/7 \lambda_T$, where $\lambda_T$ has been evaluated at the separatrix from measurements within the database. \DEdit{The total sheath heat transmission factor $\gamma$ has been fixed to 7~\cite{Stangeby_book}~\cite{Brida_2020}.} The parallel electron conductivity coefficient is given by $\kappa_e = \kappa_z \cdot 2600  $\,W\,m$^{-1}$\,eV$^{-7/2}$, where $\kappa_z = (0.672 + 0.076 \sqrt{ Z_{\mathrm{eff}} } +0.252 \cdot Z_{\mathrm{eff}} )^{-1}$ is the correction to the electron parallel thermal conductivity introduced by considering an effective ion charge $Z_{\mathrm{eff}}$~\cite{Goldston_2017}. \Edit{The assumption $Z_{\mathrm{eff}} = 2.0$ yields $\kappa_e = 2026$\,W\,m$^{-1}$ eV$^{-7/2}$ for the three devices.}\DEdit{For the three devices, $Z_{\mathrm{eff}} = 1.5$ has been assumed, giving $\kappa_e = 2275$\,W\,m$^{-1}$ eV$^{-7/2}$.}
Lastly, $f_{\mathrm{pow}}$ and $f_{\mathrm{mom}}$ have been set to 0.2 and 0, respectively, across the three devices. This choice is motivated by the fact that most of this dataset is unseeded, and, hence, momentum losses could be neglected. The obtained values for $C_{\mathrm{2pt}}$ are 6.3, 2.2 and 2.9 in C-Mod, AUG and JET, respectively. These values are very close to those found via regression, which is a remarkable result given the large number of simplifications introduced with the two-point model.

\section{Predictive formula and its applicability}
\label{sec:form}

\begin{figure}[h!]
        \centerline{\includegraphics[width=0.5 \textwidth]{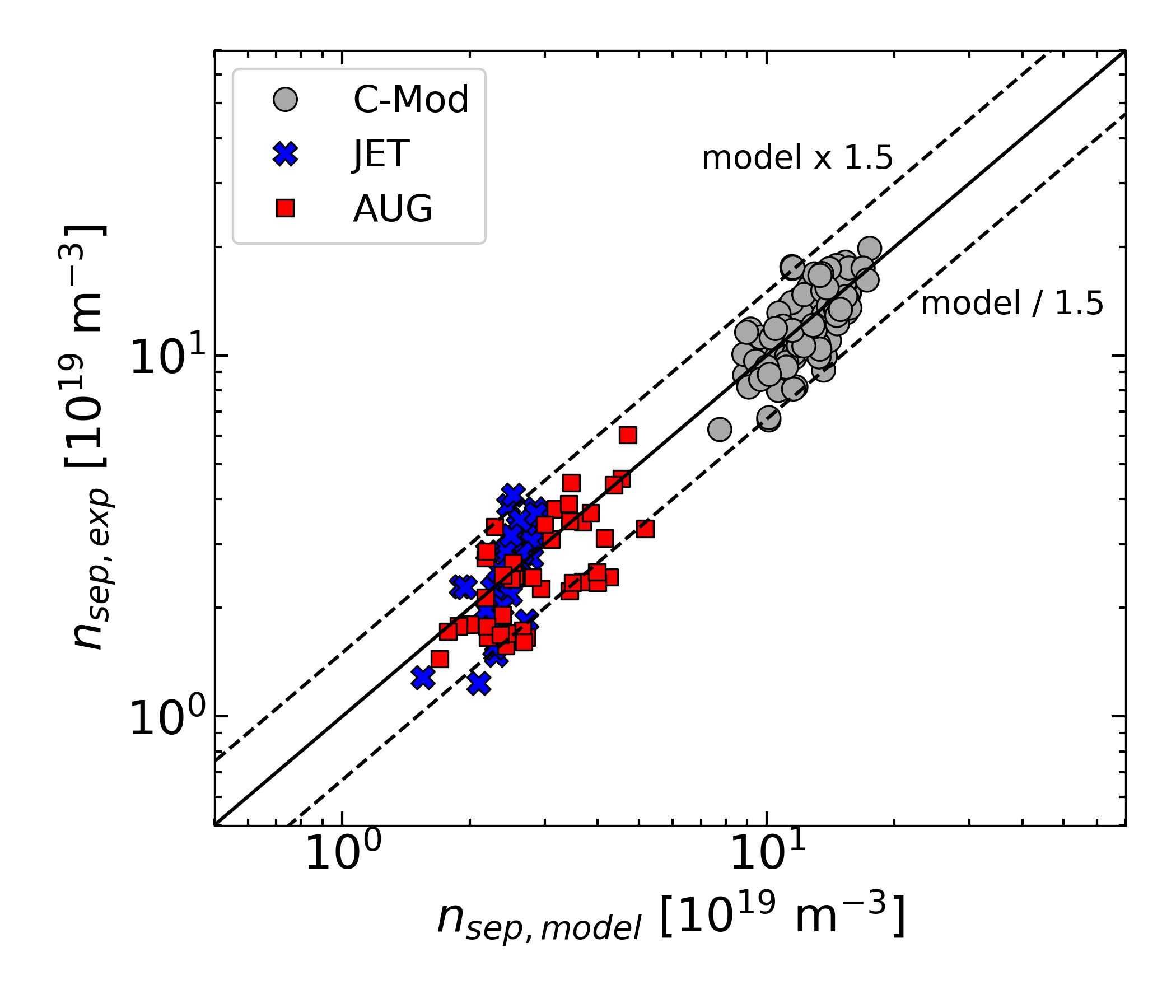}}
        \caption[]{Experimental $n_{e,\mathrm{sep}}$ values against those given by the predictive formula (Eq.~\ref{eq:nsep_mod}) for C-Mod (gray circles), AUG (red squares) and JET (blue crosses).}
        \label{nsep_model}
\end{figure}

Motivated by the overall good agreement between the regression result and the expectations from the two-point model, a fully predictive formula for $n_{e,\mathrm{sep}}$ is proposed, which reads:
\begin{equation}
    n_{e,\mathrm{sep,mod}} = C_{\mathrm{2pt}} \cdot p_{0, \mathrm{div}}^{0.20} \cdot I_{p}^{0.03} \cdot B_{t}^{-0.26} \cdot \\ 
    \left( \frac{P_{\mathrm{SOL}}}{R_{\mathrm{geo}}} \right)^{0.19} \cdot a_{\mathrm{geo}}^{-0.47},
    \label{eq:nsep_mod}
\end{equation} 
\noindent where $C_{\mathrm{2pt}}$ is given by Eq.~\ref{eq:C2pt_simple}, and all quantities are expressed in SI units, except for $I_p$ (MA), $P_{\mathrm{SOL}}/R_{\mathrm{geo}}$ (MW/m) and $n_{e,\mathrm{sep}}$ ($10^{19}$\,m$^{-3}$).
Figure~\ref{nsep_model} shows the comparison between the predicted $n_{e,\mathrm{sep}}$ values via Eq.~\ref{eq:nsep_mod} and the experimental values. The model is able to predict $n_{e,\mathrm{sep}}$ within a factor 1.5 across the three tokamaks, a level of fidelity previously unmatched in the literature.
\newline Nonetheless, many open questions on $n_{e,\mathrm{sep}}$ prediction remain. Firstly, this database encompasses data with similar closed divertor configuration, and it is known that the divertor configuration affects the relationship between $n_{e,\mathrm{sep}}$ and divertor parameters, such as $T_{\mathrm{d}}$~\cite{Lomanowski_2023, Groth_2015, Moulton_2015eps}. Future studies should investigate whether such a complex 2D geometry dependence could be captured by simple models. Secondly, most of the database used in this study is composed of unseeded plasmas, and it is well known that (1) impurity seeding is needed in a power plant to protect the plasma facing components and (2) impurity seeding lowers $n_{e,\mathrm{sep}}$~\cite{Pacher_2015, Lore_2022, Dunne_2017_hfshd}, as it reduces the available power to ionize neutrals in the SOL. From the two-point model perspective, impurity seeding will affect the divertor plasma temperature $T_{\mathrm{d}}$ and, hence, the divertor neutral pressure. This change, in turn, will impact $f_{\mathrm{pow}}$ and $f_{\mathrm{mom}}$, as they are a strong function of $T_{\mathrm{d}}$~\cite{Stangeby_2018, Lore_2022, Lipschultz_2007, Paradela_Perez_2017}. Future studies should investigate how this complex dynamic could be parametrized in a simple model. Lastly, other interesting open physics questions are how different $E \times B$ drift directions (forward vs. reversed $B_t$), wall materials (carbon vs. metal) and confinement regimes (H-mode vs. L-mode) impact $n_{e,\mathrm{sep}}$ and the proposed formula.

\section{Predictions for ITER, SPARC, DTT, JT60-SA and COMPASS-U}
\label{sec:pred}

Keeping in mind the discussion above, projections to next-step devices with similar divertor geometry may be attempted. The tokamaks considered here are ITER~\cite{Shimada_2007}, SPARC~\cite{Creely_2020}, DTT~\cite{Ambrosino_2021}, JT60-SA~\cite{Shirai_2017} and COMPASS-U~\cite{Komm_2024}, five devices expected to start operation within the next decade. Their respective parameters required to predict $n_{e,\mathrm{sep}}$ are summarized in table~\ref{table:proj}. For all the parameters, a single value is provided, except for the neutral pressure $p_{0,\mathrm{div}}$, for which a range of values is considered. This is because, for a given scenario, different gas throughput levels could be used, resulting in a range of $p_{0,\mathrm{div}}$ and, consequently, $n_{e,\mathrm{sep}}$ values. Moreover, the neutral pressure, despite being closely related to tokamak gas balance as discussed in Appendix A, is a quantity that is difficult to predict accurately without knowing the effective pumping speed of the system, which depends on the specific engineering design of the vessel. Given this uncertainty, it is more meaningful to provide a range of neutral pressures for next-step devices, rather than a single precise value.
\newline Starting with ITER, the baseline 15\,MA, Q=10 inductive H-mode burning plasma discharge is considered~\cite{Shimada_2007}. The divertor neutral pressure, $\alpha_{\mathrm{div}}$ at the outer target and $P_{\mathrm{SOL}}$ are chosen to be consistent with recent SOLPS-ITER simulations which include impurity seeding~\cite{Pshenov_2025, Pitts_2019, Pacher_2015}. Although multi-machine scalings predict $\lambda_{q,\mathrm{ITER}} \sim 1$\,mm~\cite{Eich_2013}, recent turbulent simulations found a widening of $\lambda_{q}$ at ITER-like parameters, reaching $\sim 6$\,mm~\cite{Chang_2017}. Therefore, following the argument presented in~\cite{Pitts_2019}, a value of $\lambda_q = 3.4$\,mm is adopted. To simplify the analysis, the remaining factors needed in Eq.~\ref{eq:C2pt_simple} have been kept constant across most of the tokamaks, with values $l^* = 1.2$, $\kappa_e = \Edit{2026}\DEdit{2275}$\,W\,m$^{-1}$ eV$^{-7/2}$, $f_{\mathrm{out}} = 0.65$, $C_{\mathrm{geo}} = 2.35$, $S/\lambda_q = 0.5$ (i.e. $b = 1.82$), \DEdit{$\gamma = 7$,} $B_{\mathrm{d}} / B_{\mathrm{u}} = 1.4$, $f_{\mathrm{pow}} = 0.8$ and $f_{\mathrm{mom}} = 0.5$; any deviations from these values are explicitly stated.
The last two parameters have been chosen to be consistent with a target electron temperature of $T_e = 1$\,eV, according to recent SOLPS-ITER simulations~\cite{Lore_2022}. As this ITER scenario requires a 50--50 deuterium--tritium mixture, an effective ion mass of $m_i = 2.5 \,m_{\mathrm{p}}$ is used, where $m_{\mathrm{p}}$ is the proton mass. The different fuel mass also enters the calculation of the constant $C_0$ which is equal to $ \Edit{1.35} \DEdit{0.958} \times 10^{23} $ for a 50--50 D--T mixture.
\begin{table}
\centering
\footnotesize
\begin{tabular}{l|c|c|c|c|c}
    \hline\hline
     & ITER & SPARC & DTT & JT-60SA & COMPASS-U \\
    \hline
    $I_{p}$ (MA) & 15.0 & 8.7 & 5.5 & 4.6 & 1.6 \\
    $B_{t}$ (T)  & 5.3 & 12.2 & 5.85 & 2.28 & 5.0 \\
    $P_{\mathrm{SOL}}$ (MW)  & 100 & 25 & 30 & 20 & 4.0 \\
    $R_{\mathrm{geo}}$ (m)  & 6.2 & 1.85 & 2.19 & 2.93 & 0.89 \\
    $a_{\mathrm{geo}}$ (m)  & 2.0 & 0.57 & 0.65  & 1.14 & 0.28 \\
    $p_{0,\mathrm{div}}$ (Pa)  & 2.0--20 & 1.0--10  & 1.0--10 & 0.1--1.0 & 1.0--20 \\
    $\alpha_{\mathrm{div}}$ (degree)  & 2.70 & 0.75 & 1.9 & 5.3 & 3.0 \\
    $\lambda_{q}$ (mm) & 3.4 & 0.5 & 1.5  & 1.5 & 0.8 \\
    $\hat{\kappa} $ & 1.67 & 1.64 & 1.60 & 1.60 & 1.64 \\
    $C_{\mathrm{2pt}} (10^{19}) $ & 2.6 & \Edit{7.6}\DEdit{7.5} & 3.6 & \Edit{2.3}\DEdit{2.2} & 2.5 \\
    \hline
    $n_{e,\mathrm{sep, model}}$ ($10^{19}$\,m$^{-3}$)  & \DEdit{2.5}\Edit{2.6}--4.0 & \DEdit{8.8}\Edit{8.9}--14 & 4.7--7.4 & 1.6--\Edit{2.6}\DEdit{2.5} & 4.1--\Edit{7.4}\DEdit{7.3} \\
    \hline\hline
\end{tabular}
\caption{Summary of parameter values used to estimate $n_{e,\mathrm{sep}}$ for the ITER Q=10 baseline scenario, the SPARC LSN primary reference discharge, the DTT E1 scenario, the JT-60SA baseline 4--1 scenario and the COMPASS-U high-performance H-mode scenario.}
\label{table:proj}
\end{table}
With these choices, the model predicts $n_{e,\mathrm{sep, ITER}}$ in the range \DEdit{2.5}\Edit{2.6}--4.0\,$\times 10^{19}$\,m$^{-3}$. 
\newline For SPARC, the $Q>1$ H-mode primary reference discharge in lower single null (LSN) configuration is considered~\cite{Creely_2020, Rodriguez-Fernandez_2022}. Also in this case, the divertor neutral pressure, $\alpha_{\mathrm{div}}$ at the outer target, $P_{\mathrm{SOL}}$ and $\lambda_q$ are chosen to be consistent with recent SOLPS-ITER simulations~\cite{Lore_2024, Park_2024, Kuang_2020}. Given the 50--50 D--T fuel mixture, $m_i = 2.5 \,m_{\mathrm{p}}$ and $C_0 = \DEdit{0.958}\Edit{1.35} \times 10^{23} $ are again used. The resulting $n_{e,\mathrm{sep, SPARC}}$ predictions fall in the range \DEdit{8.8}\Edit{8.9}--14\,$\times 10^{19}$\,m$^{-3}$.
\newline For DDT, the so-called LSN scenario E1 is modeled~\cite{Casiraghi_2023}, using the latest machine design at $R_{\mathrm{geo}} = 2.19$\,m~\cite{Ambrosino_2021}. The divertor neutral pressure, $\alpha_{\mathrm{div}}$ at the outer target, $P_{\mathrm{SOL}}$ and $\lambda_q$ are also chosen to be consistent with recent SOLPS-ITER and DIVGAS simulations~\cite{Moscheni_2025, Tantos_2024, Balbinot_2021}. Since this device will operate with deuterium as the main fuel, $m_i = 2 \,m_{\mathrm{p}}$ and $C_0 = \DEdit{1.07}\Edit{1.51} \times 10^{23} $ are employed. The resulting $n_{e,\mathrm{sep, DTT}}$ values predicted by Eq.~\ref{eq:nsep_mod} are in the range 4.7--7.4\,$\times 10^{19}$\,m$^{-3}$.
\begin{figure}
        \centerline{\includegraphics[width=0.5 \textwidth]{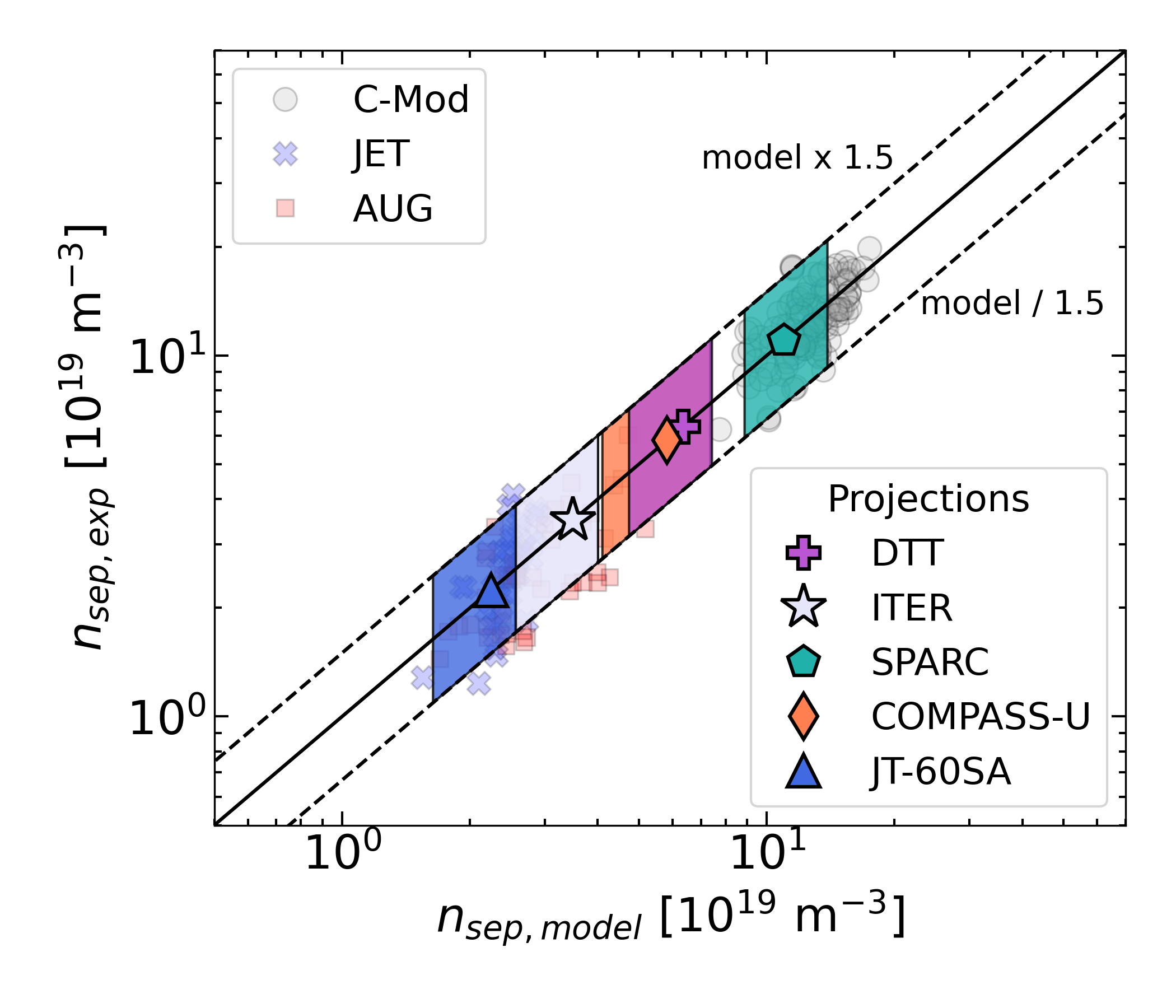}}
        \caption[]{Experimental $n_{e,\mathrm{sep}}$ values against those given by the predictive formula (Eq.~\ref{eq:nsep_mod}) for C-Mod, AUG and JET. Predictions for ITER, SPARC, DTT, JT-60SA and COMPASS-U are overlaid, including also the additional upper and lower boundaries given by the factor 1.5 uncertainty. A range of values is obtained for each device, as a range of input $p_{0,\mathrm{div}}$ is considered, see table~\ref{table:proj}. The symbol indicates one specific operational scenario at a given $p_{0,\mathrm{div}}$ (10\,Pa for ITER, 3\,Pa for SPARC, 4.5\,Pa for DTT, 0.5\,Pa for JT-60SA and 6\,Pa for COMPASS-U).}
        \label{nsep_model_pred}
\end{figure} 
\newline For JT-60SA, the baseline 4--1 scenario foreseen for the OP2 campaign, featuring ITER-like shape and $q_{95}$, is considered~\cite{JT-60SA_rp}. The chosen value of $ P_{\mathrm{SOL}} = 20$\,MW is consistent with realistic levels of installed auxiliary heating power during the OP2 campaign ($\sim 25$\,MW). The divertor neutral pressure, $\alpha_{\mathrm{div}}$ at the outer target, and $\lambda_q$ are in line with recent SOLPS-ITER~\cite{Rubino_2024eps} and SONIC simulations~\cite{Umezaki_2026}. As in the previous case, the main fuel is deuterium, and therefore $m_i = 2 \,m_{\mathrm{p}}$ and $C_0 = \DEdit{1.07}\Edit{1.51} \times 10^{23} $ are used. All other parameters are the same as those adopted for the other machines, except for $ B_{\mathrm{d}} / B_{\mathrm{u}} $ which is 1.55 in this case. With these choices, the predicted $n_{e,\mathrm{sep, JT-60SA}}$ values lie in the range 1.6--\Edit{2.6}\DEdit{2.5}\,$\times 10^{19}$\,m$^{-3}$. 
\newline Finally, for COMPASS-U, the high-performance H-mode scenario is considered~\cite{Komm_2024}. Given the similarity between COMPASS-U and C-Mod, many parameters are chosen to be close to those of C-Mod. In particular, the fraction of power going to the outer target is $f_{\mathrm{out}} = 0.5$~\cite{Terry_2013}, the neutral pressure is taken in the range 1--20\,Pa~\cite{Silvagni_2025} and $S / \lambda_q = 1$~\cite{Eich_2013}. The SOL power decay length $\lambda_q$ is obtained from Brunner’s scaling~\cite{Brunner_2018}, derived from C-Mod data. Considering deuterium as the main fuel, the predicted $n_{e,\mathrm{sep, COMPASS-U}}$ values are in the range 4.1--\Edit{7.4}\DEdit{7.3}\,$\times 10^{19}$\,m$^{-3}$.
\newline Figure~\ref{nsep_model_pred} summarizes the model predictions and includes upper and lower bounds corresponding to the factor of 1.5 uncertainty. Remarkably, despite the simplicity of the model, the $n_{e,\mathrm{sep}}$ predictions reported in table~\ref{table:proj} are consistent with values obtained from SOLPS-ITER simulations for ITER~\cite{Pshenov_2025}, SPARC~\cite{Lore_2024}, DTT~\cite{Moscheni_2025} and JT-60SA~\cite{Rubino_2024eps}. For COMPASS-U, corresponding SOLPS simulations are not currently available, therefore a comparison is not possible.
\newline This agreement allows the model to be used to interpret the observed variation of $n_{e,\mathrm{sep}}$ across the different tokamaks. The high $n_{e,\mathrm{sep}}$ observed in C-Mod and expected in SPARC is primarily driven by their large $C_{\mathrm{2pt}}$, which mainly results from the shallow grazing angle at the outer target and the narrow $\lambda_q$ characteristic of these devices. In addition, their relatively small size reduces the connection length, further increasing $n_{e,\mathrm{sep}}$. Conversely, larger devices such as ITER, JT-60SA and JET tend to exhibit lower $n_{e,\mathrm{sep}}$ values. Somewhat counter-intuitively, COMPASS-U is predicted to have lower $n_{e,\mathrm{sep}}$ than C-Mod, despite their similar size and parameters. This is due to the larger grazing angle at the outer target in COMPASS-U, which reduces $C_{\mathrm{2pt}}$ while leaving all the remaining input parameters nearly unchanged. In this respect, a direct comparison of $n_{e,\mathrm{sep}}$ between COMPASS-U and C-Mod will be particularly valuable for further validation of the model.

\section{Conclusions}
\label{sec:concl}

In this work, a multi-machine (C-Mod, AUG and JET) H-mode separatrix database is used to derive a scaling expression and to benchmark a physics-based predictive model for the separatrix electron density, $n_{e,\mathrm{sep}}$, requiring only engineering parameters as input. The database encompasses H-mode plasmas that are mainly unseeded, with similar closed divertor configuration, favorable ion $B\times \nabla B$ drift direction, and mixed ELM regimes. By analyzing this database, the following conclusions are drawn:
\begin{enumerate}
    \item The separatrix electron density shows positive dependence on the divertor neutral pressure $p_{0\mathrm{,div}}$ and $P_{\mathrm{SOL}}/R_{\mathrm{geo}}$, and negative dependence on $B_t$ and $a_{\mathrm{geo}}$, i.e. the machine size. Interestingly, $n_{e,\mathrm{sep}}$ is found to be independent of the plasma current, similarly to what has been observed in single-machine studies upon normalization to the neutral pressure dependence~\cite{Kallenbach_2018, Silvagni_2025}. Moreover, device-specific multiplication constants have been derived through regression, with the goal of comparing them to the expectation from a simple model.
    \item A theory-based expression for $n_{e,\mathrm{sep}}$ as a function of engineering parameters has been derived by coupling the two-point model equations with simple geometrical formulas. The resulting expression reproduces well both the device-specific multiplication constants and the dependencies on $P_{\mathrm{SOL}}/R_{\mathrm{geo}}$, $I_p$ and $a_{\mathrm{{geo}}}$, which is a remarkable result given the simplicity of the model. On the other hand, the exponents of $p_{0,\mathrm{div}}$ and $B_t$ differ somewhat in magnitude from the regression values, suggesting that additional hidden dependencies may be at play.
    \item Motivated by the overall good agreement between the regression results and the expectations from the two-point model, a fully predictive formula for $n_{e,\mathrm{sep}}$ is proposed. This formula combines the two-point model multiplicative constant and the regression-based dependencies on $p_{0,\mathrm{div}}$, $P_{\mathrm{SOL}}/R_{\mathrm{geo}}$, $I_p$, $B_t$ and $a_{\mathrm{{geo}}}$. The model is able to predict $n_{e,\mathrm{sep}}$ within a factor 1.5 across the three tokamaks, and provides projections to next-step devices (ITER, SPARC, DTT, JT-60SA and COMPASS-U) that are in agreement with available SOLPS simulations.
\end{enumerate}
\noindent Despite these encouraging results, this work should be regarded as a first step toward improving our predictive capabilities of $n_{e,\mathrm{sep}}$ using reduced, low-fidelity models. To make these predictions more reliable, additional important physics aspects should be taken into account. In particular, it is well known that impurity seeding and the divertor configuration have a significant impact on $n_{e,\mathrm{sep}}$~\cite{Lomanowski_2023, Groth_2015, Pacher_2015, Lore_2022}. Future studies should therefore investigate how the complex dynamic introduced by impurities and two-dimensional geometry could be captured within such reduced simple models.

\ack

The authors warmly acknowledge insightful discussion with O.J.W.F. Kardaun concerning $q_{\mathrm{cyl}}$ definitions in toroidal devices, as well as valuable exchanges with N. Aiba, T. Eich, M. Komm, P. Innocente, R. Lobel, J.D. Lore, M. Moscheni, M. Muraca, A. Pshenov, M.R.K. Wigram and E. Wolfrum. This work has been carried out within the framework of the EUROfusion Consortium, funded by the European Union via the Euratom Research and Training Programme (Grant Agreement No 101052200 — EUROfusion). Views and opinions expressed are however those of the author(s) only and do not necessarily reflect those of the European Union or the European Commission. Neither the European Union nor the European Commission can be held responsible for them. This work was supported by US DOE Awards DE-SC0014264, DE-SC0021629 and was published as part of the international project co-financed by the Polish Ministry of Science and Higher Education within the programme called `PMW' for 2023-2024.

\section*{Appendix A. Neutral pressure and tokamak gas balance}

In this appendix, additional considerations on the neutral pressure and its connection to tokamak gas balance are outlined. Considering the tokamak as a large reservoir in equilibrium, the neutral pressure measured at a certain location, $p_0$, is proportional to the gas injection rate, $\Gamma $, with the constant of proportionality given by the effective pumping speed, $S_{\mathrm{eff}}$:
\begin{equation}
    p_{0} [Pa] = \frac{k_B [J/K] \cdot T [K]}{S_{\mathrm{eff}} [m^3/s]} \cdot \Gamma [\mathrm{molecules}/s] = \frac{4.14 \times 10^{-21}}{S_{\mathrm{eff}} [m^3/s]} \cdot \Gamma [\mathrm{molecules}/s]
    \label{eq:Seff}
\end{equation}
where $k_B$ is the Boltzmann constant and $T$ the neutral temperature. The last step in Eq.~\ref{eq:Seff} has been obtained by considering that neutrals are typically at room temperature ($\sim 300$\,K) at the baratron measurement location.
This simplified picture of tokamak gas balance neglects uptake and release of gas by the wall~\cite{Loarer_2013, Rohde_2009, Lipschultz_2009}, as well as the weak dependence of pumping speed on neutral pressure.
\begin{figure}
        \centerline{\includegraphics[width=0.5 \textwidth]{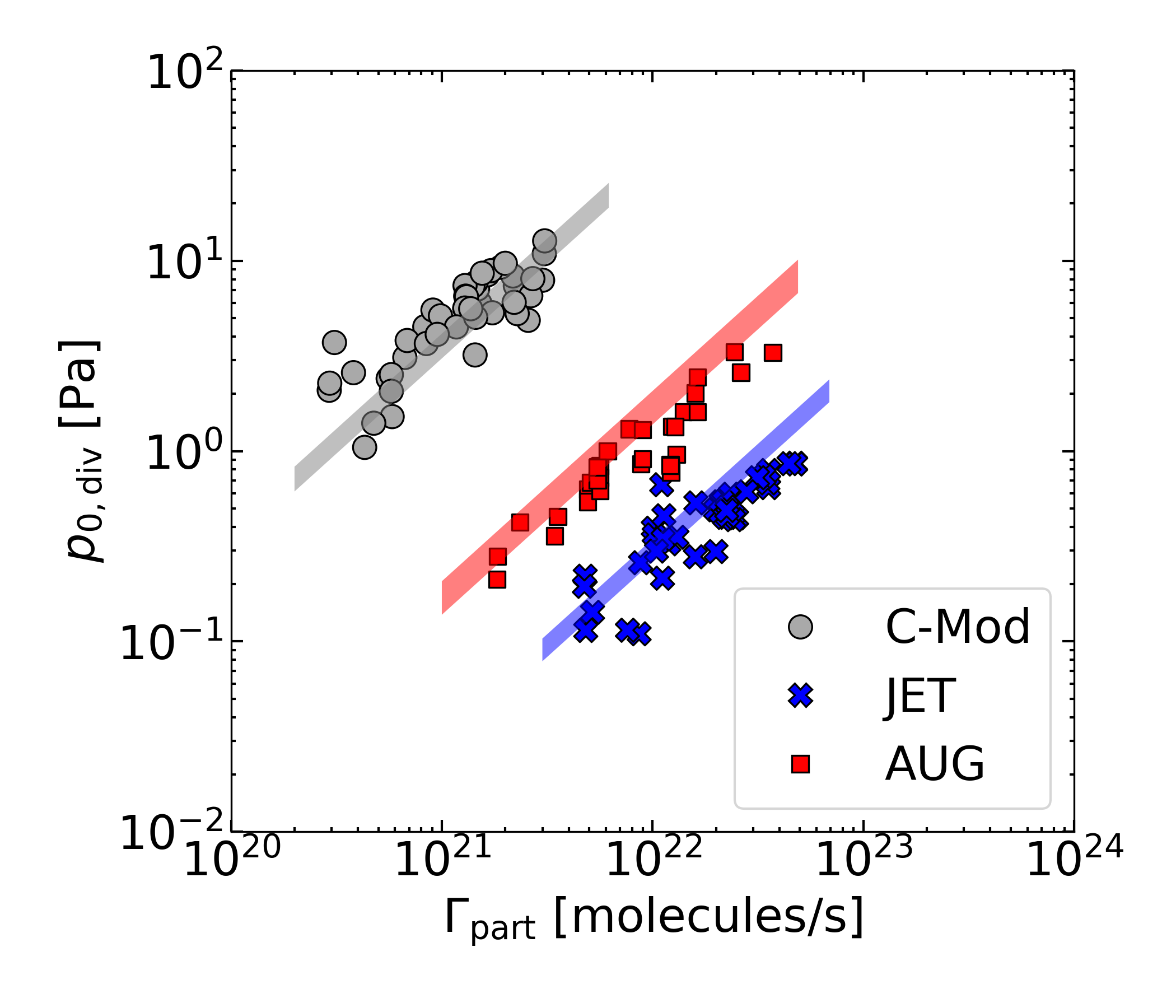}}
        \caption[]{Divertor neutral pressure against the injected particle flux for C-Mod (gray circles), AUG (red squares) and JET (blue crosses). The gray, red and blue thick lines represent the expected pressure for the measured effective pumping speed in C-Mod, AUG and JET, respectively. }
        \label{Gamma_p0div}
\end{figure} 
Figure~\ref{Gamma_p0div} shows the experimental relationship between divertor neutral pressure and injected particle flux for the three tokamaks. In AUG and JET, neutral beam injection (NBI) fueling has been neglected, as it is usually one order of magnitude lower than the gas puff. For seeded discharges in AUG, the impurity injected particle flux has also been considered. The colored thick lines indicate the expected neutral pressure for the measured effective pumping speed in the three tokamaks. Experimental data cluster around these expected pressure values, confirming---at a first approximation---the validity of this simplified tokamak gas balance picture. In AUG and JET, the effective pumping speed is set by the cryopump, and is around 20--30\,m$^3$/s in AUG~\cite{Kallenbach_2018} and 120--158\,m$^3$/s in JET~\cite{Vesa_2024eps, Vesa_master}. On the other hand, in C-Mod LSN discharges, the effective pumping speed is mainly set by the turbopumps and is expected to be in the range 1--1.35\,m$^3$/s~\cite{Irby_1988}. The upper cryopump installed in C-Mod in 2007~\cite{LaBombard_2007aps} has a negligible impact on gas balance in LSN discharges due to the low pressure in the upper divertor region (about 300 times smaller than the lower divertor pressure). Moreover, in C-Mod, the injected particle flux has been calculated by dividing the time-integrated particle content injected into the vessel by the elapsed time. This approximate calculation is necessary because H-mode C-Mod plasmas are usually run with a closed gas valve: the feedback control typically closes the valve as soon as the plasma enters H-mode to keep the line-averaged core density constant. This is a typical behavior observed in divertors with no strong particle draining driven by external pumps.
Nonetheless, it is noteworthy that, to first approximation, the simplified proportionality relation set by the effective pumping speed holds in C-Mod as well.

\section*{Appendix B. SOL power conservation}
\label{sec:appA}
In this Appendix, Eq.~\ref{eq:pow} is derived. The SOL flux-tube area perpendicular to the parallel direction $A_{\perp, \mathrm{SOL}}$ is:
\begin{equation}
    A_{\perp,\mathrm{SOL}} = 2 \pi R d_r \mathrm{sin}(\alpha) = 2 \pi R d_r \frac{B_p}{B},
\end{equation}
where $R$ is the major radius, $d_r$ is the radial width of the flux tube, $\alpha$ is the magnetic field line inclination angle and $B_p$ and $B$ are the poloidal and total magnetic field strength, respectively. Therefore, power conservation along the flux tube from the upstream `u' to the downstream `d' position reads:
\begin{gather}
    (1-f_{\mathrm{pow}}) q_{||,\mathrm{u}} R_{\mathrm{u}} d_{r,\mathrm{u}} \frac{B_{p,\mathrm{u}}}{B_\mathrm{u}}=  q_{||,\mathrm{d}} R_{\mathrm{d}} d_{r,\mathrm{d}} \frac{B_{p,\mathrm{d}}}{B_\mathrm{d}} \nonumber \\
    (1-f_{\mathrm{pow}}) q_{||,\mathrm{u}} =  q_{||,\mathrm{d}} \frac{d_{r,\mathrm{d}}}{d_{r,\mathrm{u}}} \frac{R_{\mathrm{d}}}{R_{\mathrm{u}}} \frac{B_{p,\mathrm{d}}}{B_{p,\mathrm{u}}} \frac{B_{\mathrm{u}}}{B_{\mathrm{d}}}.
    \label{eq:Pcons}
\end{gather}
We consider that the upstream radial width of the flux tube is the heat flux decay length at the outer midplane, $d_{r,\mathrm{u}}=\lambda_q$, while the target radial width is the integral heat flux decay length multiplied by the magnetic flux expansion, $d_{r,\mathrm{d}}=\lambda_{\mathrm{int}} f_{x,\mathrm{mag}}$. The magnetic flux expansion is defined as~\cite{Loarte_1999}:
\begin{equation}
    f_{x,\mathrm{mag}} = \frac{R_\mathrm{u}}{R_{\mathrm{d}}} \frac{B_{p,\mathrm{u}}}{B_{p,\mathrm{d}}}.
\end{equation}
Therefore, Eq.~\ref{eq:Pcons} becomes:
\begin{gather}
    (1-f_{\mathrm{pow}}) q_{||,\mathrm{u}} =  q_{||,\mathrm{d}} \frac{\lambda_{\mathrm{int}}}{\lambda_q} \frac{B_{\mathrm{u}}}{B_{\mathrm{d}}} \nonumber \\
    (1-f_{\mathrm{pow}}) q_{||,\mathrm{u}} =   q_{||,\mathrm{d}} b \frac{B_{\mathrm{u}}}{B_{\mathrm{d}}},
    \label{eq:Pcons2}
\end{gather}
where $b = \lambda_{\mathrm{int}} / \lambda_q = 1 + 1.64 \cdot S$ is the divertor broadening factor. Eq.~\ref{eq:Pcons2} can be intuitively understood by considering that the SOL cross-sectional area of the magnetic flux bundle is inversely proportional to the magnetic field strength (due to magnetic flux conservation under ideal MHD, see e.g.~\cite{Zohm_book}), and that the upstream heat flux is reduced by $1/b$ due to the perpendicular diffusion process taking place in the divertor chamber.

\section*{Appendix C. Cylindrical safety factor definitions for non-circular plasma cross-sections}
\label{sec:appB}

In this appendix, different definitions of the cylindrical safety factor are compared. Considering a plasma torus with circular poloidal cross-section, the edge safety factor is defined as~\cite{wesson2011tokamaks}: 
\begin{equation}
    q_{\mathrm{cyl,circ}} = \frac{a B_t}{R \langle B_p\rangle}= \frac{2 \pi}{ \mu_0} \frac{a^2 B_t }{R I_p},
    \label{eq:qcylid}
\end{equation}
with $\langle B_p\rangle = \mu_0 I_p /(2 \pi a)$. This $q_{\mathrm{cyl}}$ definition does not take into account the elongated and triangular plasma poloidal cross-sections typically used in many present-day tokamaks. Therefore, to account for the real plasma boundary contour, different versions of Eq.~\ref{eq:qcylid} have been introduced in the literature. The $q_{\mathrm{cyl}}$ definition adopted in this study follows the approach of Uckan et al.~\cite{Uckan_1990, Uckan_1991} who found that $2 \pi a_{\mathrm{geo}} \hat{\kappa}$ provides a good approximation of the plasma separatrix contour length on the poloidal plane when ITER-like (up-down symmetric) shapes are considered. Accordingly, Eq.~\ref{eq:qcyl} can be obtained from Eq.~\ref{eq:qcylid} by substituting $a$ with $a_{\mathrm{geo}} \cdot \hat{\kappa}$.
\newline A second $q_{\mathrm{cyl}}$ definition present in the literature is the one adopted by the ITPA confinement database working group, which is given by~\cite{McDonald_2007, Verdoolaege_2021}: 
\begin{equation}
    q_{\mathrm{cyl,ITPA}} = \frac{2 \pi}{ \mu_0} \frac{\kappa_a a^2_{\mathrm{geo}} B_t }{R_{\mathrm{geo}} I_p},
    \label{eq:qcylitpa}
\end{equation}
where $\kappa_{a} = V / (2 \pi R_{\mathrm{geo}} \pi a_{\mathrm{geo}}^2)$ and $V$ is the plasma volume. Additional definitions that have been proposed within the ITPA confinement working group are~\cite{Kardaun_priv}:
\begin{equation}
    q_{\mathrm{cyl,ITPA, II}} = \frac{2 \pi}{ \mu_0} \frac{(L_{S}/2\pi) a_{\mathrm{geo}}  B_t }{R_{\mathrm{geo}} I_p}
    \label{eq:qcylitpa2}
\end{equation}
\begin{equation}
    q_{\mathrm{cyl,ITPA, III}} = \frac{2 \pi}{ \mu_0} \frac{ (L/2\pi)^2 B_t }{R_{\mathrm{geo}} I_p},
    \label{eq:qcylitpa3}
\end{equation}
where $L$ represents the separatrix contour length (on the poloidal plane) evaluated from the equilibrium reconstruction, while $L_S = \frac{S}{2 \pi R_{\mathrm{geo}}}$ is an approximation of $L$ obtained from the plasma surface area $S$. Comparing the different definitions, Eq.~\ref{eq:qcylitpa3} appears to be a good generalization of Eq.~\ref{eq:qcylid} for a non-circular plasma cross-section, as it employs an effective minor radius evaluated directly from the equilibrium separatrix contour length. Nonetheless, Eq.~\ref{eq:qcylitpa3} (and Eq.~\ref{eq:qcylitpa2}) have not been adopted in the literature in view of the difficulty of assembling a multimachine database of $L$ and $L_S$.
\newline It is useful to compare the $q_{\mathrm{cyl}}$ values given by Eq.~\ref{eq:qcyl} and Eq.~\ref{eq:qcylitpa} with $q_{95}$, which is the quantity that we would like to approximate in this study. In our dataset, $q_{\mathrm{cyl,ITPA}}$ is on average 30\,$\%$ lower than $q_{95}$, whereas Eq.~\ref{eq:qcyl} yields $q_{\mathrm{cyl}}$ values that are close to $q_{95}$ within 7\,$\%$ on average. The main difference between these two definitions is the $q_{\mathrm{cyl}}$ dependence on the `effective' elongation: linear in the ITPA definition and squared in Eq.~\ref{eq:qcyl}. Since the $q_{\mathrm{cyl}}$ expression with the squared effective elongation dependence reproduces $q_{95}$ more accurately, this definition is used in this work.
\newline The curious reader might also wonder whether replacing $\hat{\kappa} $ with $ \kappa_{a} $ in Eq.~\ref{eq:qcyl} leads to any substantial difference in the $q_{\mathrm{cyl}}$ values. The effect is very small: when $\kappa_{a}$ is used, the deviation of $q_{\mathrm{cyl}}$ from $q_{95}$ is 6\,$\%$ on average, compared with a 7\,$\%$ deviation when $\hat{\kappa}$ is used.
This is due to the small discrepancy between $\hat{\kappa} $ and $ \kappa_{a}$, which is 5\,$\%$ on average.
Therefore, the interested reader may replace $\hat{\kappa}$ with $\kappa_{a}$ in the equations derived in section~\ref{sec:form} without substantially altering the result.

\section*{References}

\bibliography{cit}

\end{document}